%% file: neurips_2021.tex
\documentclass{article}

    \usepackage[preprint, nonatbib]{neurips_2021}

\usepackage[utf8]{inputenc} % allow utf-8 input
\usepackage[T1]{fontenc}    % use 8-bit T1 fonts
\usepackage[colorlinks = true,
linkcolor = black,
urlcolor = blue,
citecolor = black,
anchorcolor = black]{hyperref}       % hyperlinks
\usepackage{booktabs}       % professional-quality tables
\usepackage{amsfonts}       % blackboard math symbols
\usepackage{nicefrac}       % compact symbols for 1/2, etc.
\usepackage{microtype}      % microtypography
\usepackage{xcolor}         % colors
\usepackage{amsmath}        % allow cases in equations
\usepackage{todonotes}      % allows side todo notes
\usepackage{caption}
\usepackage{subcaption}
\usepackage{svg}            % allows svg images
\usepackage{listings}       % used to include code
\usepackage{booktabs}
\usepackage{multirow}

\title{Encoder-Decoder Architectures for Clinically Relevant Coronary Artery Segmentation}

\author{%
    João Lourenço Silva\textsuperscript{1}
    \And
    Miguel Nobre Menezes\textsuperscript{2}
    \And
    Tiago Rodrigues\textsuperscript{2}
    \And
    Beatriz Silva\textsuperscript{2}
    \And
    Fausto J. Pinto\textsuperscript{2}
    \And
    Arlindo L. Oliveira\textsuperscript{1}
    \And 
    \normalfont\textsuperscript{1}INESC-ID / Instituto Superior Técnico, University of Lisbon\\
    \nolinkurl{{joao.lourenco.silva, arlindo.oliveira}@tecnico.pt}\\
    \\
    \textsuperscript{2}Cardiology Department, CAML, CCUL, Lisbon School of Medicine, University of Lisbon\\
    \nolinkurl{{mnmenezes.gm, tiagoegrodrigues, beatrizsilvaee}@gmail.com}\\
    \nolinkurl{faustopinto@medicina.ulisboa.pt}
}

\begin{document}

\maketitle

\input{Sections/abstract}
\input{Sections/introduction}

\input{Sections/related_work}
\input{Sections/architecture}

\input{Sections/implementation}

\input{Sections/conclusion}

\input{Sections/broader_impact}

%\input{Sections/acknowledgements}

% BIBLIOGRAPHY
% Add the Bibliography to the PDF table of contents (not the document table of contents)
\pdfbookmark[0]{References}{bib}
\bibliographystyle{IEEEtran}

\bibliography{references.bib}

\appendix

\include{Sections/appendix}

\end{document}

%% file: Sections/abstract.tex
\begin{abstract}

Coronary X-ray angiography is a crucial clinical procedure for the diagnosis and treatment of coronary artery disease, which accounts for roughly 16\% of global deaths every year. However, the images acquired in these procedures have low resolution and poor contrast, making lesion detection and assessment challenging. Accurate coronary artery segmentation not only helps mitigate these problems, but also allows the extraction of relevant anatomical features for further analysis by quantitative methods. Although automated segmentation of coronary arteries has been proposed before, previous approaches have used non-optimal segmentation criteria, leading to less useful results. Most methods either segment only the major vessel, discarding important information from the remaining ones, or segment the whole coronary tree based mostly on contrast information, producing a noisy output that includes vessels that are not relevant for diagnosis. We adopt a better-suited clinical criterion and segment vessels according to their clinical relevance. Additionally, we simultaneously perform catheter segmentation, which may be useful for diagnosis due to the scale factor provided by the catheter's known diameter, and is a task that has not yet been performed with good results. To derive the optimal approach, we conducted an extensive comparative study of encoder-decoder architectures trained on a combination of focal loss and a variant of generalized dice loss. Based on the EfficientNet and the UNet++ architectures, we propose a line of efficient and high-performance segmentation models using a new decoder architecture, the EfficientUNet++, whose best-performing version achieved average dice scores of 0.8904 and 0.7526 for the artery and catheter classes, respectively, and an average generalized dice score of 0.9234. Source code will be available at: \textcolor{blue}{\url{https://github.com/jlcsilva/EfficientUNetPlusPlus}}.

%\paragraph{Keywords:} Coronary Artery Segmentation; Deep Convolutional Neural Networks; X-ray Coronary Angiography; Medical Image Segmentation.

\end{abstract}

%% file: Sections/introduction.tex
\section{Introduction}

Coronary arteries are the blood vessels that carry blood to the heart. Coronary Artery Disease (CAD), also know as Coronary Heart Disease (CHD) or Ischemic Heart Disease (IHD), is the narrowing or blockage of these arteries, caused by the build-up of atherosclerotic plaques inside them, which can lead to limited blood flow and consequent damage to the heart muscle. CAD is the cause of roughly 16\% of global deaths every year \cite{rudd2020global}.

X-ray coronary angiography (CAG) is one of the main procedures for CAD diagnosis and treatment. Patients submitted to CAG are catheterized and have their arteries filled with a radio-opaque contrast agent that makes them visible in X-ray images. Traditionally, physicians used these images to assess the presence and severity of stenosis (i.e., artery narrowing) through visual inspection. However, this method is highly subjective and potentially unreliable, which led to the development of Quantitative Coronary Angiography (QCA), a diagnostic support tool. Using semi-automatic edge-detection algorithms, QCA reports the vessel diameter at user-specified locations and the point of stenosis. Nevertheless, the low contrast and resolution of CAG images, the uneven contrast agent distribution and the presence of artifacts such as pacemakers, the spine and the catheter itself make this task very challenging. Thus, QCA still requires manual correction of the vessel boundaries before calculating the stenosis percentage, limiting its use in clinical practice. Indeed, in everyday practice, the severity of stenosis is still assessed visually in most cases, rather than with QCA software.

Recently, coronary artery segmentation performance in CAG images has been significantly improved by deep learning methods. Most of them either segment only the major vessel or try to segment the whole coronary tree based primarily on contrast differences. The procedures, however, may not be clinically optimal. The former discards potentially damaged vessels whose lesions may not be negligible, and the latter includes secondary vessels that may not be relevant for either diagnostic or therapeutic purposes, potentially distracting physicians from the important ones. We circumvent these shortcomings by adopting a better-suited clinical criterion, with the aid of expert physicians, in which a vessel is only segmented if it is important to properly assess other vessels' lesions or contains a non-negligible lesion itself. More specifically, we only segment arteries no smaller than 2 mm at their origin, as thinner vessels are generally deemed inadequate for revascularization.

Additionally, with more complex lesion assessment and anatomical feature extraction in mind, we simultaneously segment the catheter, whose known diameter provides a scale factor that may come to play an important role in current and future diagnostic methods. To the best of our knowledge, simultaneous catheter and coronary artery segmentation in CAG images has only been performed in one previous work \cite{vlontzos2018deep}, reporting dice score coefficients (DSC) of 0.54 and 0.69 for the artery and catheter classes, respectively. We obtained DSCs of 0.8904 and 0.7526 for these same classes.

Due to the adopted segmentation criteria and the inclusion of a catheter class, the problem we aim to solve is more complex than those addressed in previous work. Instead of simply learning to distinguish between vessels - or main vessels, easily identifiable by their larger width - and background, our model must also learn to segment catheters and determine the vessels' clinical relevance. To address these challenges, we adopt a loss function that combines a variant of Generalized Dice Loss (GDL) \cite{sudre2017generalised, yang2019major} and Focal Loss (FL) \cite{lin2017focal}. The former provides global segmentation quality information and is designed to handle class imbalance. The latter uses pixel-wise information to force models to focus on hard, misclassified examples, leading to improved segmentation of less common classes and challenging parts of images, such as class boundaries.

To determine the best architecture for this segmentation task, we conducted an extensive comparative study of existing encoders and decoders, which provided insights into the best architectural patterns for this and, presumably, other medical image segmentation problems. Based on these findings, we propose a new efficient and high-performance segmentation architecture, the EfficientUNet++, based on the EfficientNet family of models \cite{tan2019efficientnet} and the UNet++ \cite{zhou2018unet++} decoder architecture. Using this architecture, we achieved DSC of 0.8904 and 0.7526 for the artery and catheter classes, respectively, and a generalized dice score of 0.9234. 

Overall, the main contributions of this paper are as follows:

% 1) We propose an approach to perform simultaneous coronary artery and catheter segmentation in CAG images, using a new and better-suited clinical criterion, in which vessels are only labeled as such if they are relevant for diagnosis; 2) We perform an extensive quantitative and qualitative comparison of the performance of existing encoders and decoders, which may provide valuable insights for other medical image segmentation tasks; 3) Based on the findings of our study, we propose a line of efficient and high-performance segmentation models, based on the EfficientNet backbone and the UNet++ decoder architectures, enabling practitioners to choose a trade-off between model size and model performance, according to the available hardware and the clinical needs.

\begin{enumerate}
    \item We propose an approach to perform simultaneous coronary artery and catheter segmentation in CAG images, using a new and better-suited clinical criterion, in which vessels are only labeled as such if they are deemed relevant for diagnostic and therapeutic purposes;
    \item We perform an extensive quantitative and qualitative comparison of the performance of existing encoders and decoders, which may provide valuable insights for other medical image segmentation tasks;
    \item Based on the findings of our study, we propose a line of efficient and high-performance segmentation models based on EfficientNet backbones and the new EfficientUNet++ decoder architecture, enabling practitioners to choose a trade-off between model size and model performance, according to the available hardware and the clinical needs.
    %\item Based on the findings of our study, we propose a line of efficient and high-performance segmentation models, based on the EfficientNet backbone and the UNet++ decoder architectures, enabling practitioners to choose a trade-off between model size and model performance, according to the available hardware and the clinical needs. \todo{ver se conseguimos tempo real, ainda que com pior performance para PCI}
\end{enumerate}

%% file: Sections/related_work.tex
\section{Related Work}

\paragraph{Major vessel segmentation.}

Previous work has shown that major vessel segmentation can be improved by replacing the U-Net's encoder with popular image classification backbones, either pre-trained on ImageNet \cite{yang2019major, yang2019deep} or trained from scratch on a relatively small dataset composed of 3200 CAG images \cite{xian2020main}. Additionally, it has also been shown that the use of a modified generalized dice loss function that uses weights to offset class imbalance and features a tunable penalty for false positives and false negatives could further improve segmentation performance \cite{yang2019major}. In the sequence of these findings, we train our models using a combination of the proposed loss function and focal loss and compare their segmentation quality when using different state-of-the-art encoders. 

Other authors have proposed a U-Net-based nested encoder-decoder architecture, named T-Net \cite{jun2020t}. To simplify the optimization process, the authors replaced the U-Net's blocks with residual ones. In addition, to enable feature reuse, they arranged the pooling and up-sampling operations to make all the feature maps extracted by the encoder available to every layer of equal or greater depth of the decoder, in a DenseNet-like fashion. These modifications enhanced information flow through the network and enabled it to outperform a standard U-Net. The use of dense connections is also present in the U-Net++ \cite{zhou2018unet++}, which we explore in our work. 

\paragraph{Whole vessel body segmentation.}

One of the main challenges of the coronary artery segmentation task is the discrimination between vessels and artifacts. Given that arteries are only visible in the presence of contrast, previous work has used the images acquired before contrast injection as a second-channel input to help a U-Net discern between the vessels and the background \cite{fan2018multichannel}. However, for this approach to be effective, it must be coupled with an image alignment algorithm to compensate for the motion caused by heartbeat and respiration. Furthermore, it requires the entire angiographic sequence to be acquired with minimal table motion, which can be hard to achieve, as standard clinical practice involves moving the patient table to follow the flow of dye within the vessels. For this reason, we do not use this technique in our work. 

In line with what we propose in this paper, some authors have also attempted to use different segmentation architectures to achieve better performance than what is possible with the commonly used U-Net. Specifically, they proposed using a pre-trained PSPNet \cite{zhao2017pyramid} and a U-Net++ combined with a feature pyramid network to improve multi-scale feature detection \cite{zhao2020semantic}. Other proposals include a fully convolutional encoder-decoder network specifically designed for vessel segmentation, featuring Gaussian convolutions and trained with deep supervision \cite{samuel2021vssc}.

Other authors have proposed to use a multi-layer perceptron to produce segmentation masks based on the multi-scale features extracted by multi-scale Gaussian matched and Gabor filters \cite{santhi2021automated, cervantes2019automatic}, with one of them achieving state-of-the-art performance on the coronary tree segmentation task \cite{santhi2021automated}. However, this method does not suit our needs, as the task we aim to solve requires the use of higher-level features that make it possible to discern between catheter and vessels and determine the clinical relevance of each vessel. More specifically, we only segment arteries no smaller than 2 mm at their origin, as thinner vessels are generally deemed inadequate for revascularization.

A number of proposals \cite{liang2020coronary, wang2020coronary, hao2020sequential} have explored the use of temporal information to complement the 2D data of the frame to be segmented. Some have adopted simple approaches such as replacing the input blocks of 2D segmentation models with 3D blocks capable of processing inputs composed of multiple frames \cite{wang2020coronary} or using a temporal filter to compute the weighted average of successive output frames to mitigate the presence of artifacts and inter-frame flicker. More complex alternatives have also been studied, such as the use of a U-Net-based architecture with a 3D encoder and a 2D decoder \cite{hao2020sequential}. Regardless of the strategy adopted, the use of temporal information led to quantitative improvements in segmentation performance. Nevertheless, it has not been experimentally confirmed that the obtained results are clinically more relevant. Due to the uneven distribution of contrast during the angiographic sequence and the motion caused by respiration and heartbeat, it is usual for some parts of the vessels and lesions to be more visible in some frames than others. Thus, temporal context may help mitigate false positives caused by the presence of artifacts and false negatives induced by hard to perceive vessels. However, there is also a non-negligible risk that it may lead some lesions to go unnoticed if the models attribute more importance to frames in which lesions are not perceivable than those in which they are evident. Due to this concern, we consider that the use of temporal context should be subject to in-depth analysis and leave its study for future work.

Orthogonal work has featured a coarse-to-fine approach in which the segmentation mask produced by a Fully Convolutional Network (FCN) is sparsely enhanced by the use of a region-based segmentation model \cite{thuy2021coronary}. Other authors have used a dual-path sliding-window CNN to combine local information and global context from two different sized patches to produce a vessel probability map, which they then refined using a similar network and an edge map computed from the angiogram \cite{nasr2018segmentation}. Even though both methods perform well on the coronary tree segmentation task, their reliance on local information makes them unsuitable for our task, which requires global context to determine the clinical relevance of each vessel.

To address the scarcity of labeled data and alleviate the burden on annotators, previous work has introduced a framework for weakly supervised learning that allows generic segmentation models to learn from pseudo labels generated by automatic vessel enhancement \cite{zhang2020weakly}. The model is trained using a self-paced scheme, in which it learns from pixels in descending difficulty order. In each iteration, according to the combined uncertainty of the model and the vessel enhancement algorithm for each part of the image, local online manual annotation refinement is performed. This procedure allows the pseudo labels to be improved and the model to learn well without as much effort from the annotators as in fully supervised methods. We do not use weakly nor semi-supervised learning in this work, but it could be a future line of research, as it addresses label scarcity, one of the main issues in medical image segmentation tasks.

\paragraph{Catheter and coronary tree segmentation.}

To the best of our knowledge, simultaneous catheter and coronary tree segmentation has only been addressed in a single previous work, in which a U-Net-based Siamese architecture was trained with automatically generated annotations \cite{vlontzos2018deep}. Using noisy low-level binary segmentation and optical flow, the authors generated multi-class annotations that were successively improved in a multistage segmentation approach. While this task is similar to ours, in the sense that it performs both catheter and artery segmentation, it does not address the clinical relevance issue and aims to segment every vessel of the coronary tree.

\paragraph{Other segmentation criteria.}

To the best of our knowledge, an intermediate segmentation criterion, in which secondary arteries with diameter inferior to 1 mm at their origin are not segmented, has been proposed only once \cite{iyer2021angionet}. In our work, we use a similar criterion but only segment vessels no smaller than 2 mm at their origin, as thinner vessels are generally deemed inadequate for revascularization. To perform this task, the authors couple an Angiographic Processing Network (APN), trained to learn the best possible preprocessing filter to improve segmentation performance, with the U-Net and DeepLabV3+ \cite{chen2018encoder} architectures. The results show that the APN module helps both models to achieve better results. While the APN and preprocessing, in general, can improve segmentation performance in multiple tasks, including the one we are addressing, in this paper we focus on the segmentation network's architecture and leave the study of preprocessing methods to future work.

%% file: Sections/architecture.tex
\section{Architecture}

Given the profusion of high-performing segmentation models, we decided to conduct an extensive comparative study of existing encoders and decoders to determine the best architectures for this task and what makes them perform better than the others. In this section, we describe the metrics used for model comparison and the experiments we conducted. As a result of our study, we propose a new efficient and high-performing architecture, the EfficientUNet++, based on the EfficientNet image classification models and the UNet++ decoder.

\subsection{Evaluation Metrics}\label{sec:eval_metrics}

We evaluate the segmentation quality of each class using DSC, precision and recall. The overall segmentation quality of each image is evaluated using DSC and generalized dice score (GDS). 

Let C be the number of classes, N be the number of pixels, G be the one-hot encoded ground-truth, with $g_{ci} \in \{0, 1\}$ denoting whether pixel i belongs to class c or not, and P be the predicted probabilistic map, with $p_{ci} \in [0, 1]$ representing the probability of pixel i belonging to class c. Then, %GDS takes the form:

\begin{equation}\label{eq:gds}
    GDS = 2 \frac{\sum_{c=1}^C w_c \sum_{i=1}^N g_{ci} p_{ci}}{\sum_{c=1}^C w_c \sum_{i=1}^N g_{ci} + p_{ci}},
\end{equation}

where $w_c$ is the weight assigned to class c. When all weights are set to 1, GDS is equivalent to DSC. By setting all weights to $w_c = 1 / (\sum_{i=1}^N g_{ci})^2$, each class's contribution to the score is corrected by the inverse of its area, reducing the correlation between region size and dice score \cite{crum2006generalized}. Consequently, GDS attributes the same importance to all classes independently of their frequency, making it a fairer metric for multi-class segmentation performance than DSC.  

\subsection{Loss Function}\label{sec:loss_function}

The problem we aim to solve can be interpreted as the combination of two sub-tasks: a macro-level and a micro-level one. The former consists in identifying the vessels and catheters, distinguishing them from each other and from artifacts, and determining which arteries are clinically relevant. The latter concerns the precise delineation of class contours, which is crucial for a reliable diagnosis based on the produced segmentation masks.

To address these tasks, we propose the use of a loss function composed of a variant of generalized dice loss \cite{sudre2017generalised, yang2019major}, pGDL, which provides information on global segmentation quality, and focal loss \cite{lin2017focal}, FL, which provides a pixel-wise evaluation focused on the harder pixels, which are usually the ones belonging to less common classes, near class boundaries and in artifact regions. The loss function used can be defined as:

\begin{equation}
    Loss = pGDL + \lambda FL,
\end{equation}

where $\lambda$ is a hyperparameter that controls the weight given to each term. For simplicity, we attribute the same importance to both terms and use $\lambda = 1$ in all our experiments. 

\paragraph{Generalized Dice Loss.}

As its name suggests, GDL is an extension of the dice loss (DL) function. By attributing weights to the segmentation classes, it provides invariance to different label set properties. Using the notation defined in Section \ref{sec:eval_metrics}, GDL takes the form:

\begin{equation}
    GDL = 1 - GDS = 1 - 2 \frac{\sum_{c=1}^C w_c \sum_{i=1}^N g_{ci} p_{ci}}{\sum_{c=1}^C w_c \sum_{i=1}^N g_{ci} + p_{ci}},
\end{equation}

where GDS is the function defined in Eq. \ref{eq:gds}. When all class weights, $w_c$, are set to 1, GDL is equivalent to DL. We adopt an even more generalized version of the DL function, pGDL, which adds a penalty for false positives and false negatives to GDL \cite{yang2019major}. It is defined by:

\begin{equation}
    pGDL = \frac{GDL}{1 + k(1-GDL)},
\end{equation}

where k is a hyperparameter that controls the weight of this additional penalty. When k is set to 0, pGDL is equivalent to GDL. In this work, we set k to 0.75, as we empirically verified that this value worked well for all models and led to better performance than $k = 0$, for most of them. Concerning the class weights, we follow Sudre et al. \cite{sudre2017generalised} and set them to $w_c = 1 / (\sum_{i=1}^N g_{ci})^2$. This weighting corrects the contribution of each class by the inverse of its area \cite{crum2006generalized}, reducing the correlation between region size and dice score and consequently mitigating the negative effects of class imbalance. 

\paragraph{Focal Loss.}

Focal loss is based on the cross-entropy (CE) loss function and was originally designed for one-stage object detection scenarios with extreme imbalance between foreground and background classes \cite{lin2017focal}. For notational convenience and based on the notation above, we define:

\begin{equation}
    p'_{ci} =
    \begin{cases}
        p_{ci} & \text{if $g_{ci} = 1 $} \\
        1 - p_{ci} & \text{if $g_{ci} = 0$}
    \end{cases}
\end{equation} 

and rewrite the CE loss function as $\text{CE}(p'_{ci}) = -log(p'_{ci})$. Thus, FL takes the form:

\begin{equation}
    FL(p'_{ci}) = -\alpha' (1 - p'_{ci})^\gamma log(p'_{ci}),
\end{equation}

where $\alpha \in [0, 1]$ is a weighting factor that balances positive and negative examples, with $\alpha'$ defined analogously to $p'_{ci}$, and $(1 - p'_{ci})^\gamma$ is a modulating factor, controlled by $\gamma \geq 0$, that down-weights easy examples and forces the model to focus on and learn from hard ones. When $\alpha' = 1$ and $\gamma = 0$, FL is equivalent to CE loss. As $\gamma$ increases, so does the importance given to hard examples compared to easy ones. For simplicity, we followed the FL function's authors \cite{lin2017focal} and used $\gamma = 2$ and $\alpha = 0.25$, as these values allowed to obtain good results. %Despite having obtained good results with these values, fine-tuning would probably improve performance.

% , as these values allowed us to obtain good results. However, these parameters were originally chosen for an object detection scenario. Thus, fine-tuning them would probably lead to even better performance.

\subsection{Encoder Comparison}\label{sec:encoder_comparison}

Previous work \cite{yang2019major, yang2019deep, xian2020main} has shown that the U-Net's performance can be enhanced by replacing its encoder with more sophisticated image classification architectures, both when using transfer learning from a large dataset, like ImageNet \cite{yang2019major,yang2019deep}, and when training from scratch on a relatively small dataset composed of 3200 XRA images \cite{xian2020main}, suggesting that the U-Net's original backbone is too rudimentary and not an as good feature extractor as those designed for image classification. 

On this assumption, we trained multiple models to perform our segmentation task, using image classification architectures pre-trained on ImageNet as encoders. To avoid overfitting the encoders to our small dataset and to evaluate the quality of the visual representations learned from ImageNet, their weights were frozen during decoder training, with the additional benefit of shortening the networks' training time. Furthermore, to investigate the existence of synergies between certain encoder-decoder pairs, we trained each backbone with multiple decoders. In particular, we used the U-Net \cite{ronneberger2015u}, commonly used for medical image segmentation, the UNet++ \cite{zhou2018unet++}, which has been shown by its authors to outperform the U-Net in multiple medical image segmentation tasks, and the DeepLabV3+ \cite{chen2018encoder}, a state-of-the-art semantic segmentation architecture.

\begin{figure}
    \centering
    \begin{subfigure}[b]{0.49\linewidth}
        \centering
        \includegraphics[width=\linewidth]{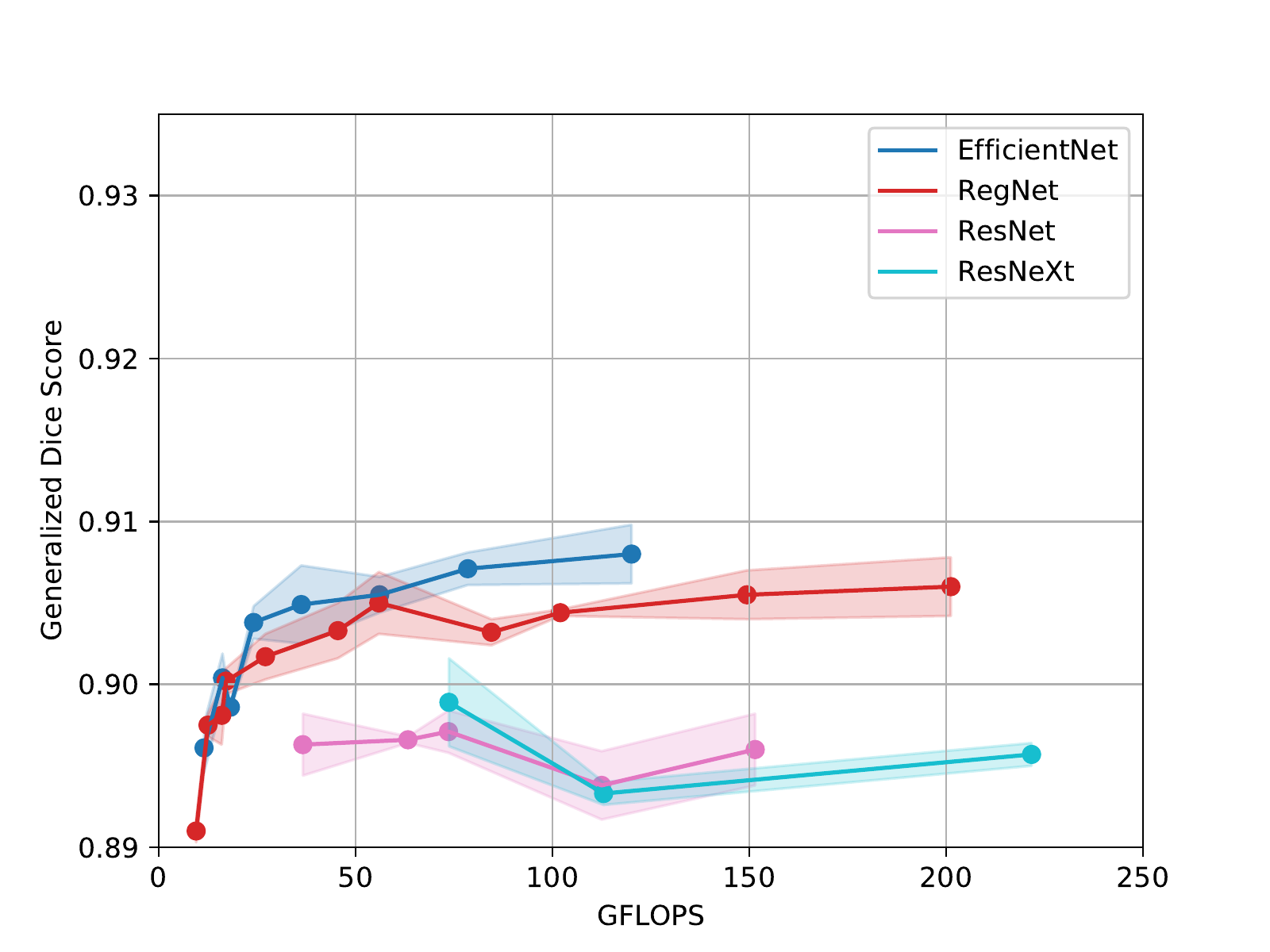}
        \caption{DeepLabV3+ decoder}
        \label{fig:deeplabv3+gflops}
    \end{subfigure}
    \begin{subfigure}[b]{0.49\linewidth}
        \centering
        \includegraphics[width=\linewidth]{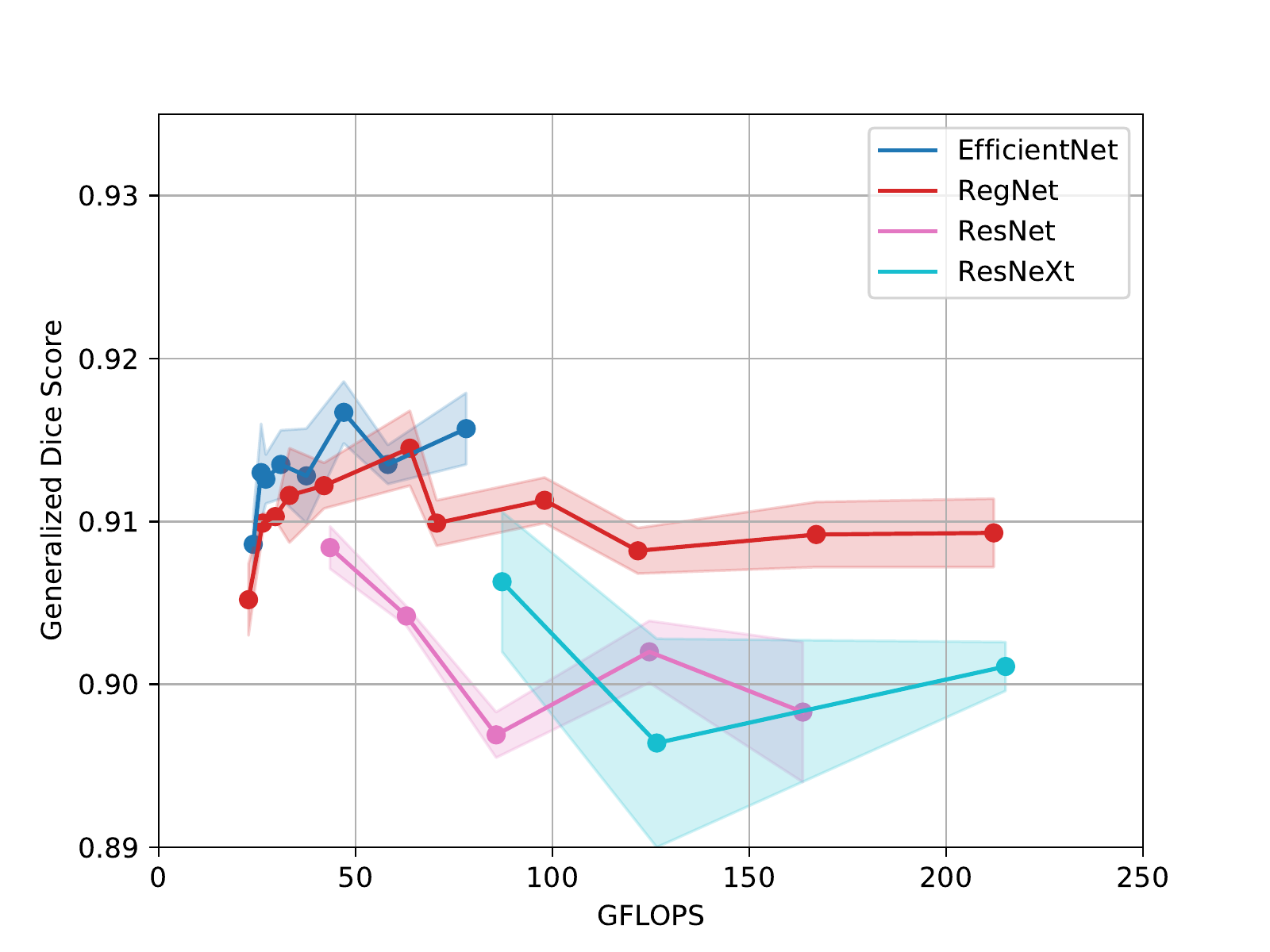}
        \caption{U-Net decoder}
        \label{fig:unetgflops}
    \end{subfigure}
    \begin{subfigure}[b]{0.49\linewidth}
        \centering
        \includegraphics[width=\linewidth]{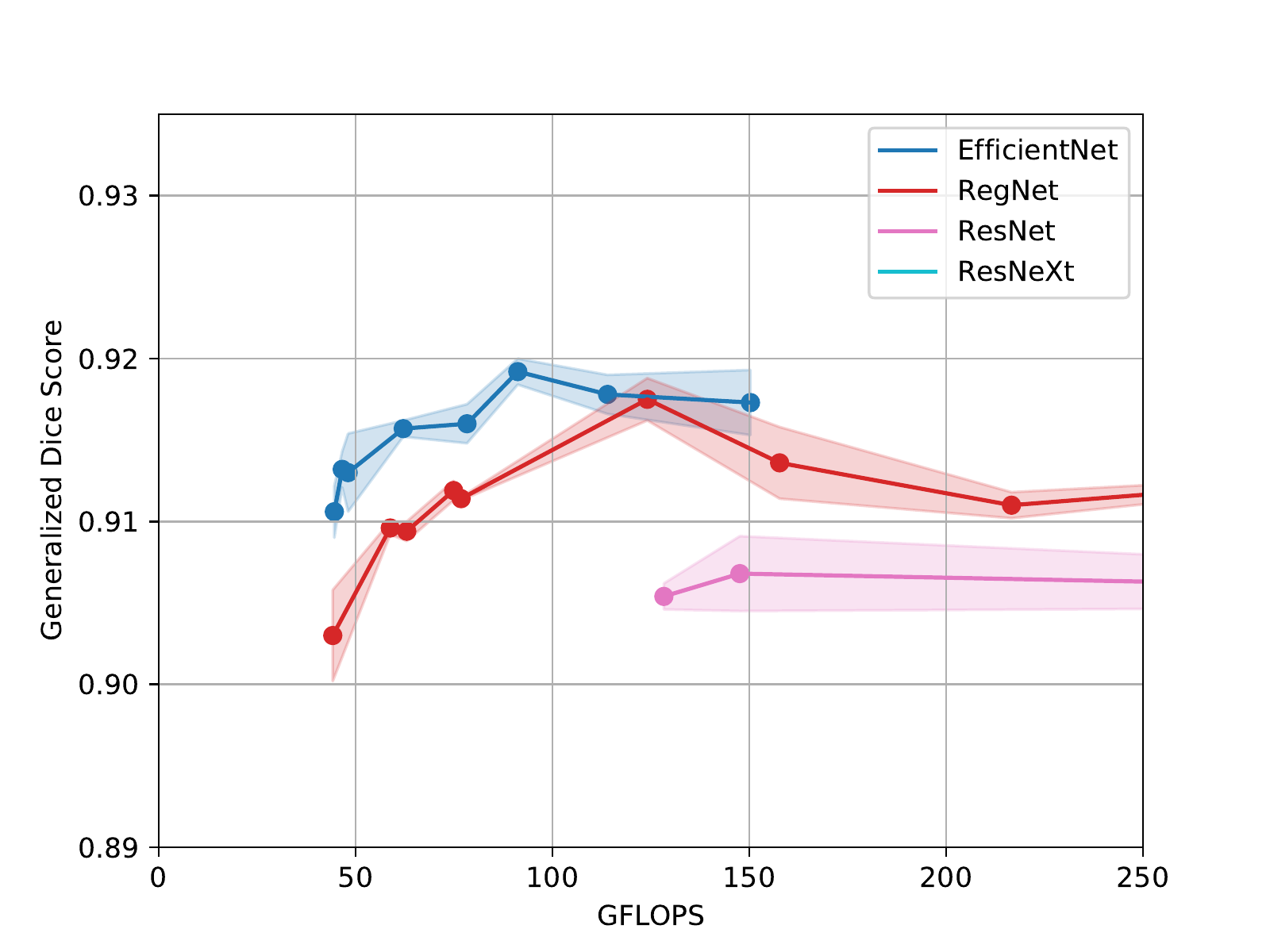}
        \caption{UNet++ decoder}
        \label{fig:unet++gflops}
    \end{subfigure}
    \begin{subfigure}[b]{0.49\linewidth}
        \centering
        \includegraphics[width=\linewidth]{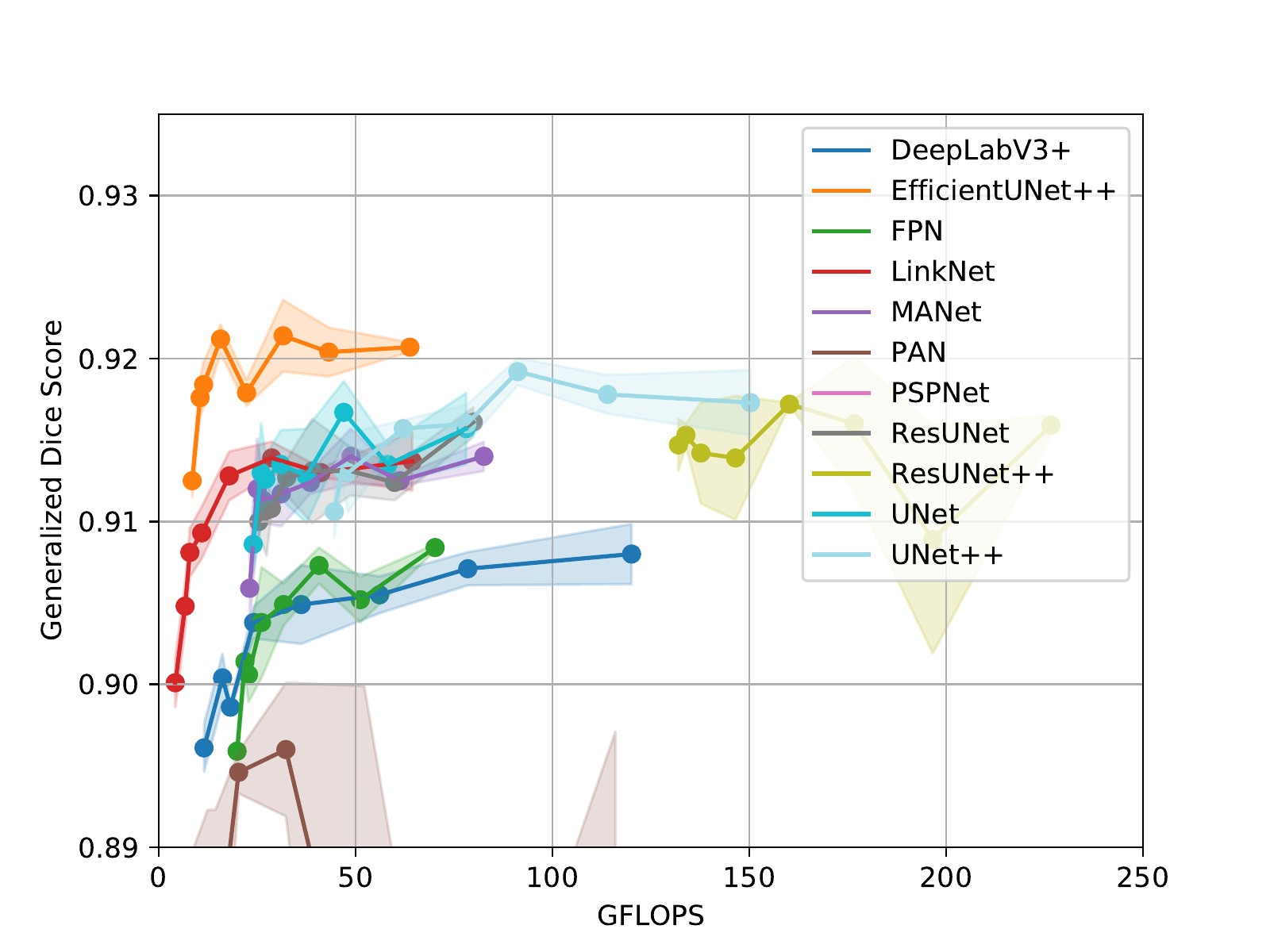}
        \caption{EfficientNet B0 to B7 encoders}
        \label{fig:decoder_gflops}
    \end{subfigure}
    \caption{Segmentation performance, measured by generalized dice score (GDS), as a function of the number of FLOPS. Figures (a), (b) and (c) show the performance of different encoders combined with the (a) DeepLabV3+, (b) U-Net and (c) UNet++ decoders. Figure (d) shows the performance of different decoders combined with the EfficientNet B0 to B7 encoders. Each polygonal line corresponds to an encoder family. The dots represent the following models, in ascending order of FLOPS: EfficientNet - B0, B1, B2, B3, B4, B5, B6, B7; RegNet - Y2, Y4, Y6, Y8, Y16, Y32, Y40, Y64, Y80, Y120, Y160; ResNet - 18, 34, 50, 101, 152; ResNeXt - 50\_32x4d, 101\_32x4d, 101\_32x8d. Above 250 GFLOPS, performance keeps degrading and is omitted. Models with GDS below 0.89 are also omitted.}
    \label{fig:comparison_results}
\end{figure}

Figures \ref{fig:deeplabv3+gflops}, \ref{fig:unetgflops} and \ref{fig:unet++gflops} show the measured segmentation performances as a function of the total number of FLOPS, when using encoders from the RegNetY \cite{radosavovic2020designing}, ResNeXt \cite{xie2017aggregated} and ResNet \cite{he2016deep} families. Graphs of performance as a function of the number of parameters are included in the appendix. Notably, for every decoder, the best performance is achieved using an EfficientNet backbone. Furthermore, for the same performance, EfficientNet encoders are always more efficient than other backbones, parameter-wise but especially computation-wise. Due to their compound scaling, EfficientNets are generally thinner than other encoders at each scale, i.e., use fewer channels to represent information. Thus, as decoder computation scales with feature map dimension, EfficientNets allow building much more computationally efficient systems than wider encoders. This is particularly evident when using complex decoders, such as the UNet++, that heavily process extracted features.

We observe that, in general, better image classification architectures allow higher segmentation performance, leading us to the intuitive and widely accepted premise that image classification performance is correlated with feature extraction capabilities. However, we notice that, for some decoders and encoder families, segmentation performance starts degrading after a certain encoder complexity is surpassed. As the encoder weights are not updated, and the same decoders converge to better solutions when using simpler encoders of the same family, we conclude that the source of degradation is decoder overfitting. Degradation is especially noticeable when using the UNet++ decoder. We hypothesize that, combined with their heavy processing of extracted features, their high number of parameters leads them to overfit the training set. The same happens for the U-Net, but not as noticeably, as it has fewer parameters and performs less processing of extracted features. Decoder overfitting could probably be mitigated or even solved using regularization techniques, a larger dataset than the one we use, composed of 270 CAG, or both. Interestingly, EfficientNet encoders seem to have a regularizing effect on decoders, suggesting they generalize and transfer better to new tasks than other models, which can be related with their thinner feature maps.

Besides all these advantages, the low number of channels used by EfficientNets to represent extracted features improves memory efficiency, allowing relatively larger training batches, which can be very important for researchers with limited hardware resources.

\subsection{Decoder Comparison}\label{sec:decoder_comparison}

Given the EfficientNet models' superior performance and efficiency, we used them as encoders for decoder comparison experiments, with each decoder being trained using all EfficientNet backbones. Figure \ref{fig:decoder_gflops} shows the segmentation performance as a function of the number of FLOPS for the DeepLabV3+ \cite{chen2018encoder}, FPN \cite{lin2017feature}, LinkNet \cite{chaurasia2017linknet}, MANet \cite{li2020multi}, PAN \cite{li2018pyramid}, PSPNet \cite{zhao2017pyramid}, ResUNet \cite{zhang2018road}, ResUNet++ \cite{jha2019resunet++}, U-Net \cite{ronneberger2015u} and UNet++ \cite{zhou2018unet++} decoder architectures. Graphs of performance as a function of the number of parameters are included in the appendix.

The results of our experiments confirm the importance of the skip connections between encoder and decoder featured in U-Net-based models. The UNet++ and ResUNet++ achieve the best performances among all models, and the LinkNet, MANet and ResUNet achieve good results, similar to the U-Net's. However, the PAN decoder also uses skip connections but has poor performance, suggesting that the attention mechanisms it uses in its blocks and bridge module are prejudicial for this task. 

The role played by attention mechanisms is not very clear. While they seem to be what harms the PAN's performance, in the ResUNet++ they appear to be beneficial, and in the MANet they do not seem to have any effect, as it performs very similarly to the U-Net in which it is based. The importance of residual connections is also unclear, as they reduce the ResUNet's performance compared to the U-Net but work well in the ResUNet++. Given that attention mechanisms and residual connections alone do not seem to improve performance in the MANet and ResUNet, respectively, we hypothesize that it is the combination of both that allows the ResUNet++ to perform so well. However, confirming this theory would require an in-depth comparison of the attention mechanisms used in the ResUNet++, MANet and PAN, and their synergies with residual connections, which we leave for future work.

Interestingly, the UNet++, which performs similarly to the ResUNet++ but more efficiently, both parameter and computation-wise, does not use residual connections nor attention. Instead, it uses densely connected nested decoder sub-networks, which promote feature reuse and allow it to extract more information from the encoder's feature maps at each scale.

Architectures based on pyramid pooling and feature pyramids are the worst-performing ones. Due to the lack of skip connections to provide accurate localization information, these models produce coarser segmentation maps than those of the U-Net-based models. While this allows them to perform well on generic segmentation tasks, it harms their performance when applied to medical images, which require fine segmentation. This is the case of the DeepLabV3+, which uses atrous convolutions and an atrous spatial pyramid pooling (ASPP) module in the encoder, the FPN, whose segmentation is based on feature pyramids, and the PSPNet, which obtains local information by applying pyramid pooling to the output of an encoder with dilated convolutions. 

\subsection{EfficientUNet++ Architecture}

When coupled with EfficientNet backbones, the UNet++ achieves high segmentation performance with reasonable parameter and computational efficiency. However, while the number of parameters is not a major concern, the computation required for inference can be prohibitive of widespread clinical usage, as it requires expensive hardware to be run promptly for entire angiographic sequences, usually comprised of about a hundred frames.

To address this, we propose a new architecture, the EfficientUNet++, that reduces computational complexity by replacing the UNet++'s blocks with residual bottlenecks with depthwise convolutions. Furthermore, to enhance performance, we process the bottleneck feature maps with concurrent spatial and channel squeeze and excitation (scSE) blocks \cite{roy2018concurrent}, which combine the channel attention of squeeze and excitation (SE) blocks \cite{hu2018squeeze} with spatial attention.

As shown in Figure \ref{fig:decoder_gflops}, when combined with EfficientNet encoders, our decoder architecture establishes a line of efficient and high-performance segmentation models, which obtained a maximum average generalized dice score of 0.9239 when using the EfficientNetB5 backbone. For the artery and catheter classes, this model obtained average dice scores of 0.8904 and 0.7526, respectively.

% Future work could include the use of network architecture search (NAS) and a systematic approach, like the one used by the RegNet authors, to derive a better and even more efficient decoder architecture. For example, in our experiments, we do not explore the number of convolution groups, and always depthwise convolutions (number of groups equal to the number of channels). However, using a smaller number of groups could potentially improve segmentation performance without much harm to efficiency. Other attention mechanisms could also prove useful, and the UNet++ could possibly be trained with deep supervision and pruned at inference time, as suggested by its authors.

\begin{figure}
    \centering
    \begin{subfigure}[b]{0.24\linewidth}
        \centering
        \includegraphics[width=\linewidth]{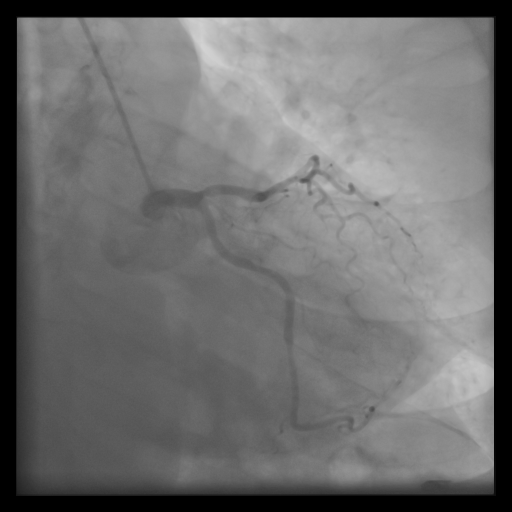}
        \caption{Angiogram}
        \label{fig:angiogram}
    \end{subfigure}
    \begin{subfigure}[b]{0.24\linewidth}
        \centering
        \includegraphics[width=\linewidth]{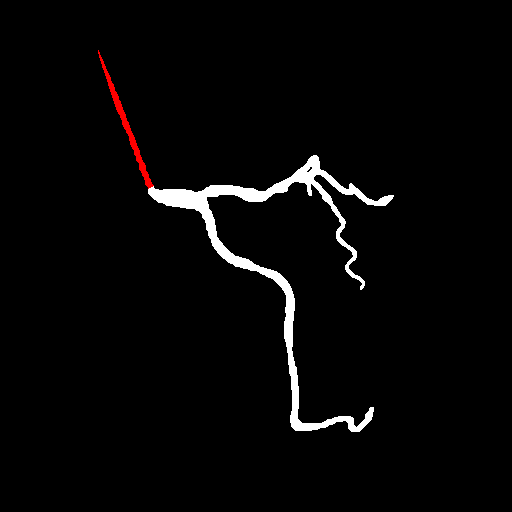}
        \caption{Ground-truth (GT)}
        \label{fig:gt}
    \end{subfigure}
    \begin{subfigure}[b]{0.24\linewidth}
        \centering
        \includegraphics[width=\linewidth]{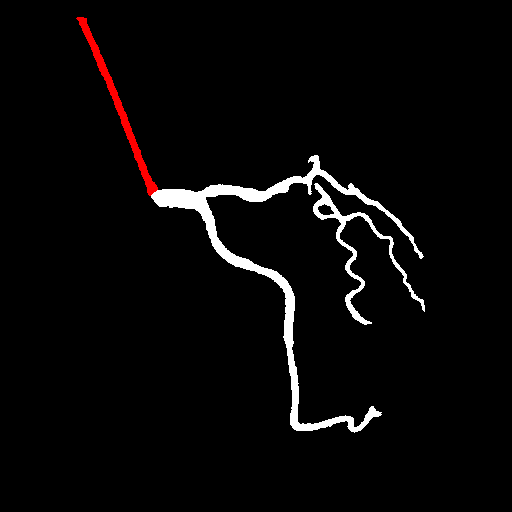}
        \caption{EfficientUNet++}
        \label{fig:segmentation}
    \end{subfigure}
    \begin{subfigure}[b]{0.24\linewidth}
        \centering
        \includegraphics[width=\linewidth]{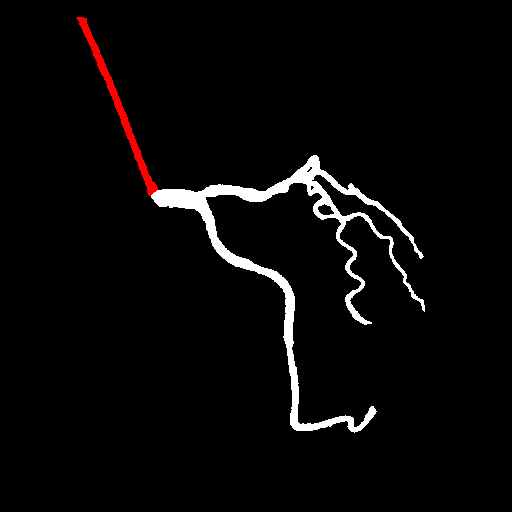}
        \caption{Enhanced GT}
        \label{fig:enhanced_gt}
    \end{subfigure}
    \caption{In some cases, the segmentation model produces better segmentation masks than humans. Here, the automatic segmentation mask, (c), was used by physicians to enhance the original ground-truth, (b), and produce the mask shown in (d).}
    \label{fig:qualitative_comparison}
\end{figure}

\begin{figure}[h!]
    \centering
    \includegraphics[width=\linewidth]{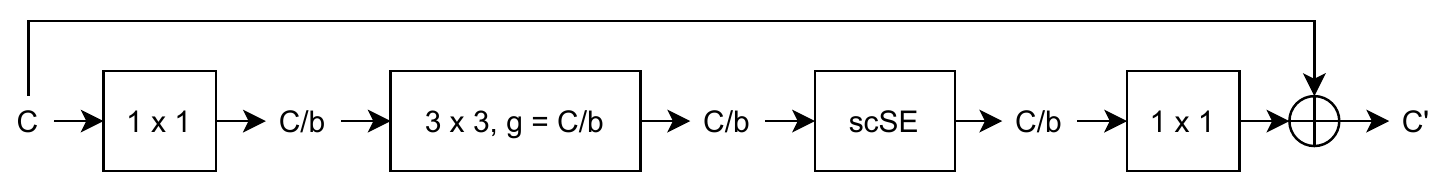}
    \caption{EfficientUNet++'s convolutional block. Each convolution is followed by batch normalization \cite{ioffe2015batch} and Hardswish \cite{howard2019searching}, except for the last $1 \times 1$ convolution, which is not activated. C and C' are the numbers of input and output channels. Feature map height and width are not altered. We set the bottleneck ratio, b, to 1, and the number of convolution groups, g, to be equal to the number of input channels, making the $3 \times 3$ convolution depthwise. The scSE block uses a squeeze ratio of 1.}
    \label{fig:block}
\end{figure}

%% file: Sections/implementation.tex
\section{Implementation Details}\label{sec:implementation_details}

\paragraph{Training details.}\label{sec:training_details} We used encoders pre-trained on ImageNet and froze their weights during decoder training. Kaiming \cite{he2015delving} and Xavier initialization \cite{glorot2010understanding} were used to initialize decoder weights in hidden and output layers, respectively. The models were trained on 4 Tesla V100S GPUs for 150 epochs each, using Adam with $\beta_1 = 0.9$, $\beta_2 = 0.999$, a mini-batch size of 8, no weight decay, and an initial learning rate of 0.001. The learning rate was divided by 10 at the $50^{th}$ and $100^{th}$ epochs. 

In most experiments, we used public PyTorch implementations \cite{Yakubovskiy:2019} under an MIT license. To keep comparisons fair, we extended the repository with ResUNet, ResUNet++ and EfficientUNet++ implementations. To obtain average scores and standard deviations, each model was trained and tested three times.

\paragraph{Dataset.} 

Our dataset is exclusively composed of retrospective data, whose use was approved by the ethics committee. It comprises 270 anonymized $512 \times 512$ pixels CAG images annotated by three expert cardiologists. The images were acquired from multiple viewing angles of the left (LCA) and right coronary arteries (RCA) of 47 patients. The dataset was split into a training and a test set, composed of 237 and 33 images, respectively. The images of each set were carefully chosen to keep them representative of the original one, having approximately identical distributions regarding the observed arteries, viewing angles and number of images annotated by each physician. In total, the training set and test set contain 165/72 and 23/10 images of the LCA/RCA, respectively.

%Our dataset is composed of 270 anonymized CAG images annotated by three expert cardiologists. All images have a resolution of 512 by 512 pixels and were acquired from multiple viewing angles of the left (LCA) and right coronary arteries (RCA) of 47 patients. The dataset was split into a training set and a test set, composed of 237 and 33 images, respectively. These sets contain 165/72 and 23/10 images of the LCA/RCA. The images of each set were carefully chosen to keep both of them representative of the original one, having approximately identical distributions regarding the observed arteries, viewing angles and number of images annotated by each physician. In total, there are 165/72 and 23/10 images of the LCA/RCA in the training and test set, respectively. In this work, we exclusively use retrospective data, with the approval of the ethics committee.

\paragraph{Data augmentation.} Our image augmentation policy consists of the sequential application of the following transformations: 1) rotation of the image at a random angle between $\pm 20$º, as this is the approximate imaging angle range; 2) horizontal and vertical shifts at random rates within $\pm 10\%$; 3) zoom at a random rate between -10\% and 10\%; 4) brightness change at a random ratio within $\pm 40\%$, to account for the variability of the angiography images' brightness across acquisition devices. 

Data augmentation is performed online. Each time an image is sampled in a batch, three augmentations are created, increasing the number of images seen per epoch by a factor of 4. Due to the near-infinite number of possible distinct augmentations, it is very unlikely for the model to see the same augmentation more than once. Thus, the number of different images seen by the model rises by a factor equal to the number of training epochs compared to offline data augmentation.

%Data augmentation is performed online. Each time an image is sampled in a batch, three augmentations are created, increasing by four times the number of images seen per epoch. Due to the near-infinite number of possible distinct augmentations, it is very unlikely for the model to see the same augmentation more than once. Thus, the number of different images seen by the model rises by a factor equal to the number of training epochs, when compared to offline data augmentation. 

%% file: Sections/conclusion.tex
\section{Discussion and Future Work}

In this work, we propose a new and better-suited clinical criterion for simultaneous catheter and artery segmentation in CAG images. Whereas most previous approaches either segment only the major vessel or the whole coronary tree, based mostly on contrast information, we segment arteries according to their clinical relevance, as assessed by expert physicians. To derive the best approach for this task, we compared multiple encoder and decoder architectures, which we trained on a combination of focal loss and a variant of generalized dice loss, which aggregates information on global and pixel-wise segmentation quality, handles class imbalance, and forces models to focus on hard pixels in class boundary and artifact regions. 

Among the existing architectures, we found the EfficientNet and the UNet++ to be the best encoder and decoder, respectively. Due to their compound scaling, EfficientNets are efficient not only in terms of parameters and computation, but also in the way they represent features, generally using fewer channels at each scale than other models, with the latter's benefit being threefold: 1) it requires less computation from decoders; 2) it seems to have a regularizing effect; 3) it slightly reduces memory use during training. Based on the EfficientNet and the UNet++, we propose a new decoder architecture, the EfficientUNet++. By replacing the UNet++'s blocks residual bottlenecks with depthwise convolutions and using scSE spatial and channel attention blocks, our architecture outperforms all other we tested and is much more computationally efficient than the UNet++. 

This paper gives origin to multiple directions of future work. Regarding model architecture, it would be interesting to study the impacts of different attention mechanisms in segmentation performance. The findings of such research could then be used to build models even better than the EfficientUNet++. Another promising line of research is the application of semi-supervised learning to this and other medical image segmentation tasks to take advantage of the great amount of existing unlabeled data.

%% file: Sections/broader_impact.tex
\section*{Broader Impact}

%In this paper, we present a new and better clinically-suited criterion for segmentation of coronary X-ray angiography images and prove that current state-of-the-art encoder and decoder architectures are able to achieve high levels of performance in this task.

This paper presents a new and better-suited clinical criterion for segmentation of coronary X-ray angiography images and proves that current state-of-the-art encoder and decoder architectures can achieve high levels of performance in this task. With these findings, we hope to open the way for a new line of research on clinically relevant coronary artery segmentation, which has not yet been explored and whose results may be of great usefulness for clinical practice.

Additionally, to derive the best approach for this task, we conducted an extensive comparative study of existing encoder and decoder architectures, whose findings will hopefully be useful for other medical image segmentation tasks. As a results of this study, we propose a new segmentation architecture, the EfficientUNet++, that combines the EfficientNet encoders with a more efficient and better-performing version of the UNet++ decoder to build a line of efficient and high-performance segmentation models. By providing all the code and implementation details, we hope other researchers and engineers will further improve it and use it in their segmentation tasks. 

Having been designed to be used as a support diagnostic tool, any possible failure of this system will be safeguarded by existing diagnostic methods and will not put patients' lives at risk.

Our method does not leverage biases in the data and, to the best of our knowledge, no one will be put at disadvantage by the results of this research.

%% file: Sections/appendix.tex
\section{Performance vs. Computation Trade-Off}

\begin{figure}[h!]
    \centering
    \includegraphics[width=0.48\linewidth]{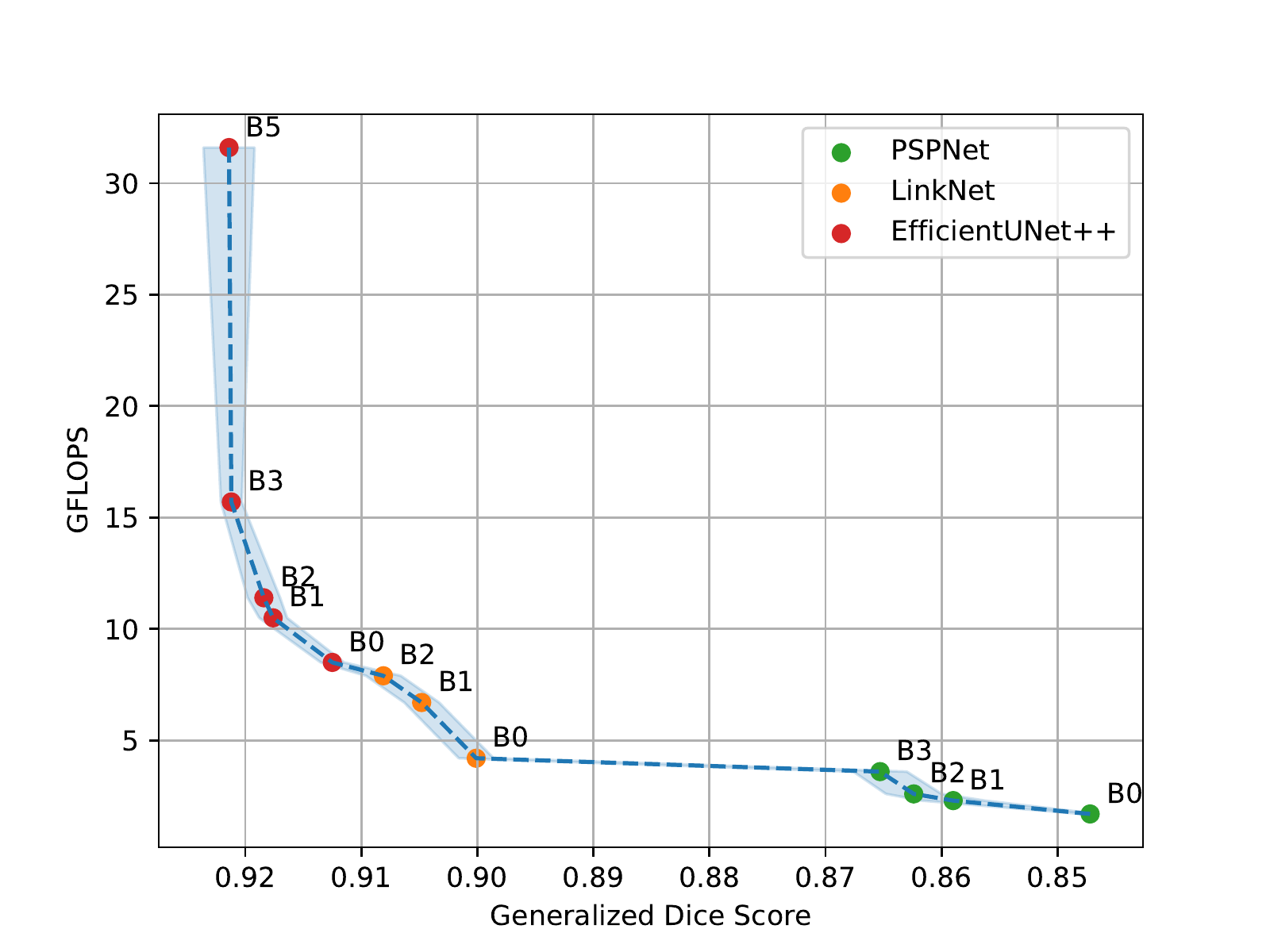}
    \caption{FLOPS as a function of performance, measured by generalized dice score. The dashed polygonal line corresponds to the Pareto frontier. Each dot represents a model: the text labels indicate the EfficientNet backbone, from B0 to B7, and the colors denote the decoder architecture.}
    \label{fig:flops_pareto}
\end{figure}

In clinical practice, depending on the available hardware resources and clinical needs, it may be necessary to make a trade-off between performance and computational efficiency. To help practitioners make that choice, we present the Pareto frontier of all tested models in Figure \ref{fig:flops_pareto}, with the number of FLOPS as a function of performance, measured by generalized dice score. Using these axes, each model on the Pareto frontier is the most efficient at each performance level, and best-performing at each computation regime. Thus, all other models need not be considered when making a trade-off between performance and computational efficiency.

In Figure \ref{fig:flops_pareto}, the lower a point is, the less computation it requires for inference, and the more to the left it is, the better its performance. Therefore, the gentler the slope between a model A and a better-performing model B, the more significant is the relative merit of A compared to B. Taking that and the low performance of PSPNet decoders into account, we suggest using the slightly more computationally demanding LinkNet-based architectures when there is a limit on computational complexity. 

\clearpage

\section{Attention Blocks Ablation}

\begin{figure}[h!]
    \centering
    \begin{subfigure}[b]{0.48\linewidth}
        \centering
        \includegraphics[width=\linewidth]{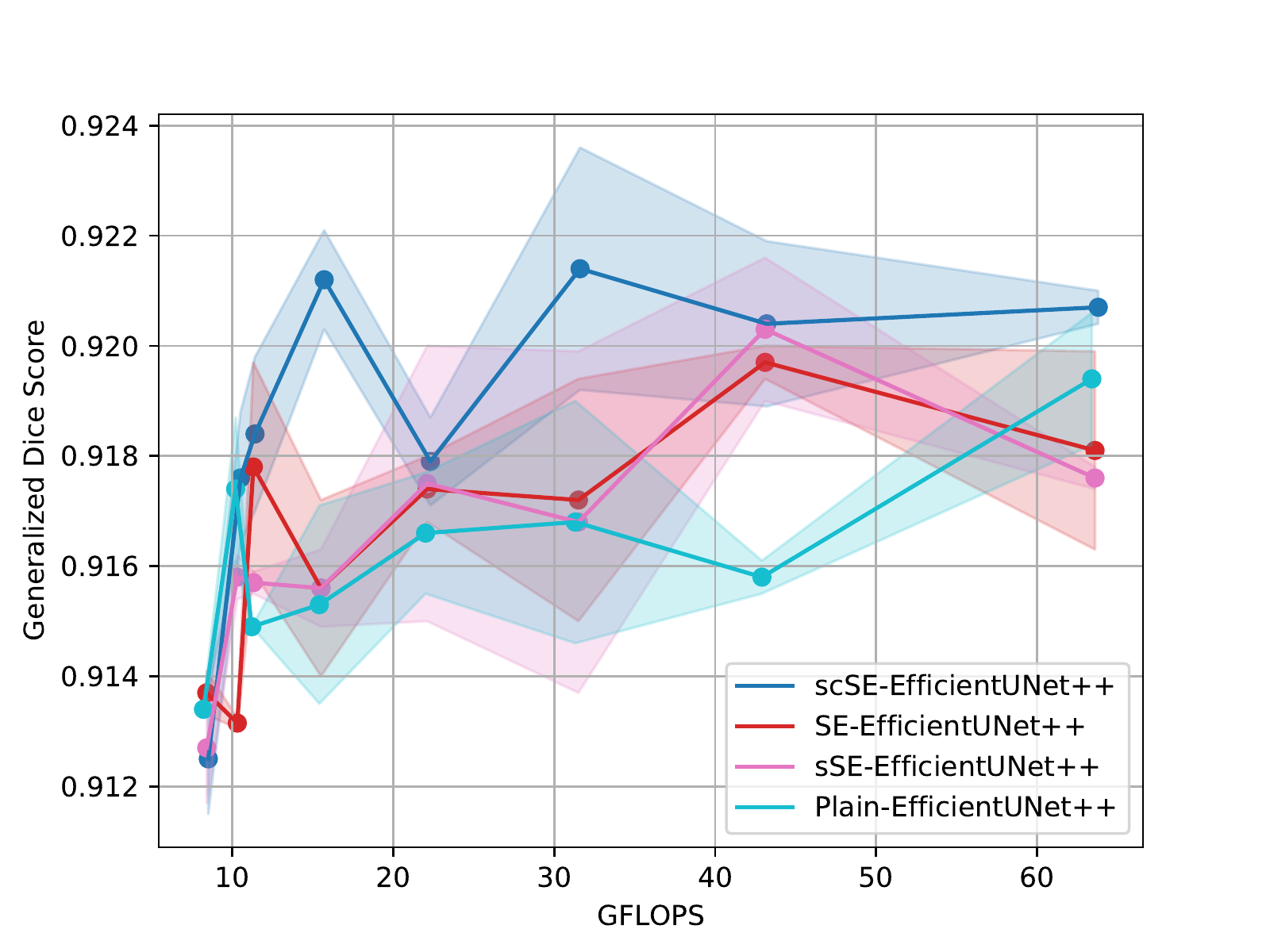}
        \caption{GDS as a function of the number FLOPS}
        \label{fig:se_vs_scse_gflops}
    \end{subfigure}
    \begin{subfigure}[b]{0.48\linewidth}
        \centering
        \includegraphics[width=\linewidth]{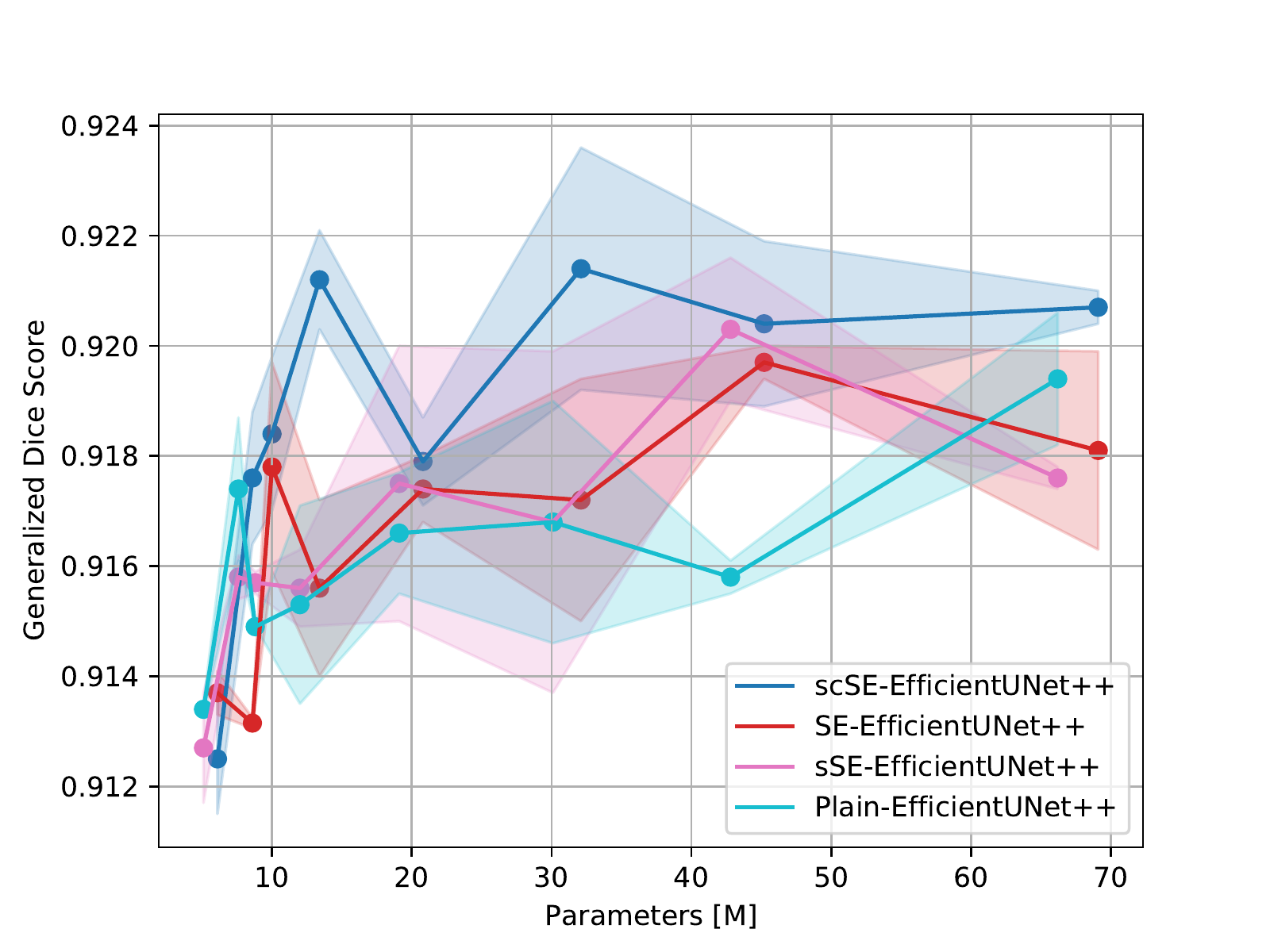}
        \caption{GDS as a function of the number of parameters}
        \label{fig:se_vs_scse_params}
    \end{subfigure}
    \caption{EfficientUNet++ performance without attention mechanisms (in cyan) and when using sSE spatial attention blocks (in pink), SE channel attention blocks (in red), and scSE concurrent channel and spatial attention blocks (in blue). The dots represent the following EfficientNet models, in ascending order of FLOPS and parameters: B0, B1, B2, B3, B4, B5, B6, B7.}
    \label{fig:se_vs_scse}
\end{figure}

Figures \ref{fig:se_vs_scse_gflops} and \ref{fig:se_vs_scse_params} show the performance of the EfficientUNet++ decoder architecture without attention mechanisms and when using SE, sSE and scSE attention blocks, which perform channel attention, spatial attention, and concurrent channel and spatial attention. From the graph, we conclude that, for high computation regimes, combining channel and spatial attention improves performance at the cost of only a slight increase in computation and parameters. At low computation regimes, all models have similar performance and computational efficiency. Thus, scSE blocks are an overall better choice, and we use them regardless of the computation regime.

Notably, even without attention mechanisms, the best EfficientUNet++ and UNet++ models perform similarly, showing that replacing the UNet++'s blocks with residual bottlenecks with depthwise convolutions increases efficiency without performance loss. 

We hypothesize that, as EfficientUNet++ models using scSE blocks generally outperform those using SE blocks, architecture variants using other attention mechanisms could exceed the performance of both, possibly with less computation and fewer parameters. Therefore, we argue that the study of attention in decoder architectures may be a promising line of future research.

\clearpage

\section{Qualitative Decoder Comparison}

\begin{figure}[h!]
    \centering
    \begin{subfigure}[b]{0.24\linewidth}
        \centering
        \includegraphics[width=\linewidth]{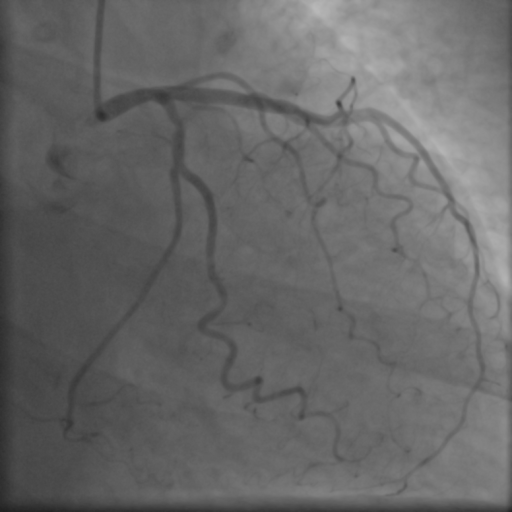}
        \caption{Angiogram}
        \label{fig:lca_img}
    \end{subfigure}
    \begin{subfigure}[b]{0.24\linewidth}
        \centering
        \includegraphics[width=\linewidth]{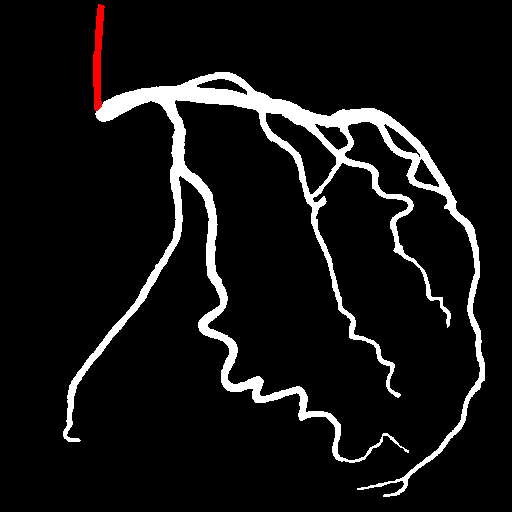}
        \caption{Ground-truth}
        \label{fig:lca_mask}
    \end{subfigure}
    \begin{subfigure}[b]{0.24\linewidth}
        \centering
        \includegraphics[width=\linewidth]{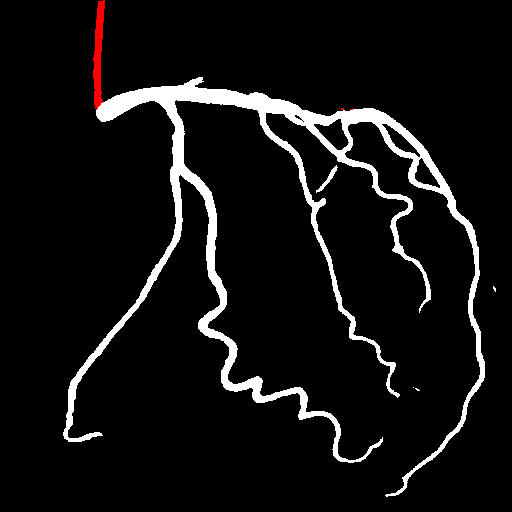}
        \caption{EfficientUNet++}
        \label{fig:lca_efficientunet++}
    \end{subfigure}
    \begin{subfigure}[b]{0.24\linewidth}
        \centering
        \includegraphics[width=\linewidth]{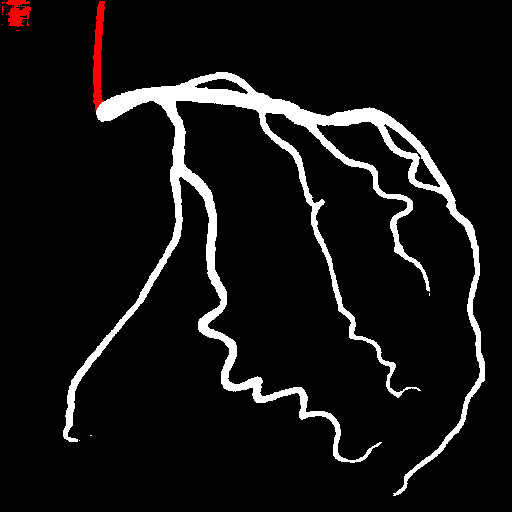}
        \caption{ResUNet++}
        \label{fig:lca_resunet++}
    \end{subfigure}
    \begin{subfigure}[b]{0.24\linewidth}
        \centering
        \includegraphics[width=\linewidth]{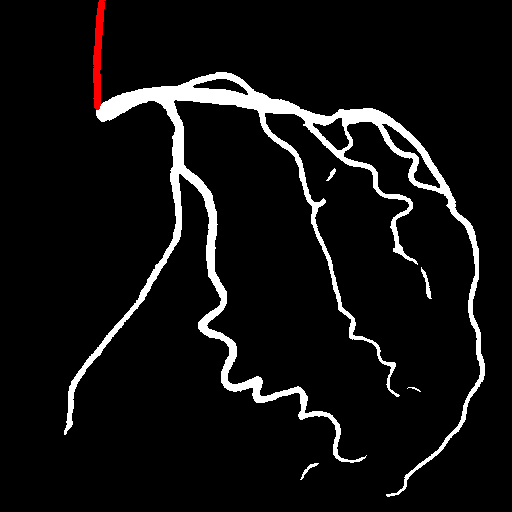}
        \caption{UNet++}
        \label{fig:lca_unet++}
    \end{subfigure}
    \begin{subfigure}[b]{0.24\linewidth}
        \centering
        \includegraphics[width=\linewidth]{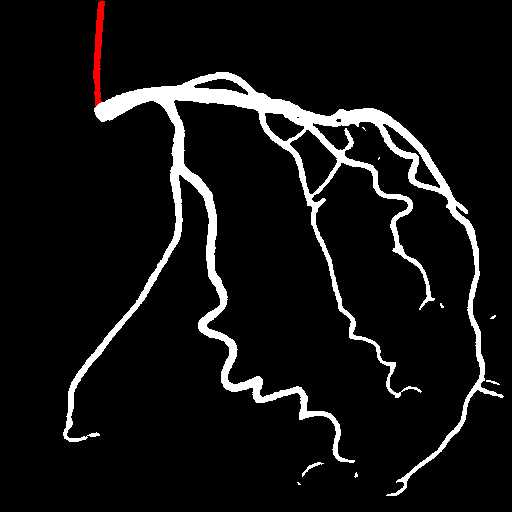}
        \caption{U-Net}
        \label{fig:lca_unet}
    \end{subfigure}
    \begin{subfigure}[b]{0.24\linewidth}
        \centering
        \includegraphics[width=\linewidth]{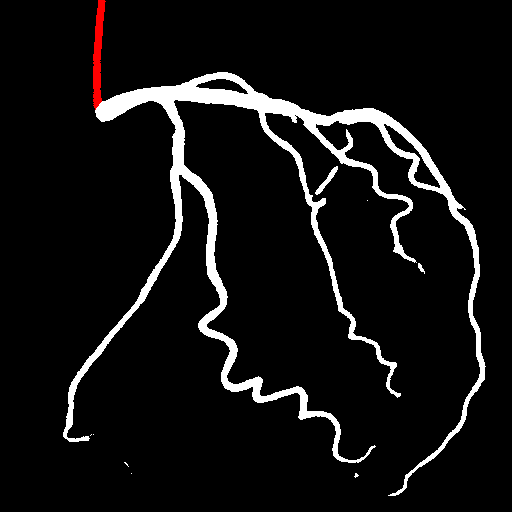}
        \caption{ResUNet}
        \label{fig:lca_resunet}
    \end{subfigure}
    \begin{subfigure}[b]{0.24\linewidth}
        \centering
        \includegraphics[width=\linewidth]{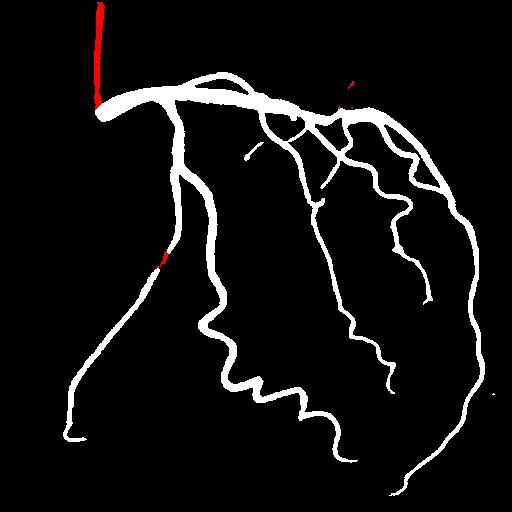}
        \caption{MANet}
        \label{fig:lca_manet}
    \end{subfigure}
    \begin{subfigure}[b]{0.24\linewidth}
        \centering
        \includegraphics[width=\linewidth]{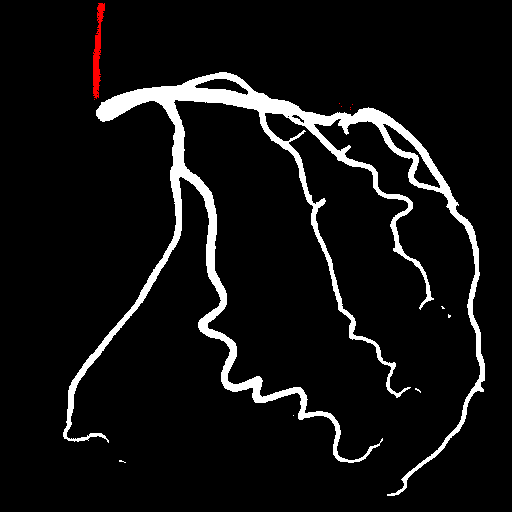}
        \caption{LinkNet}
        \label{fig:lca_linknet}
    \end{subfigure}
    \begin{subfigure}[b]{0.24\linewidth}
        \centering
        \includegraphics[width=\linewidth]{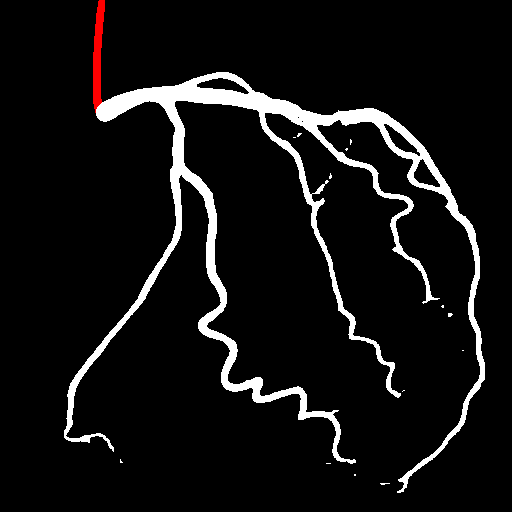}
        \caption{FPN}
        \label{fig:lca_fpn}
    \end{subfigure}
    \begin{subfigure}[b]{0.24\linewidth}
        \centering
        \includegraphics[width=\linewidth]{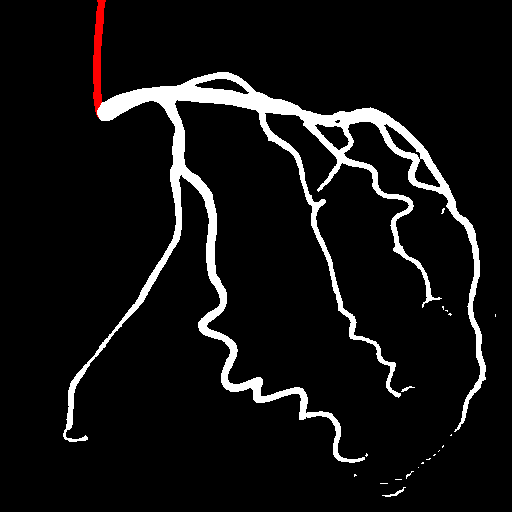}
        \caption{DeepLabV3+}
        \label{fig:lca_deeplabv3+}
    \end{subfigure}
    \begin{subfigure}[b]{0.24\linewidth}
        \centering
        \includegraphics[width=\linewidth]{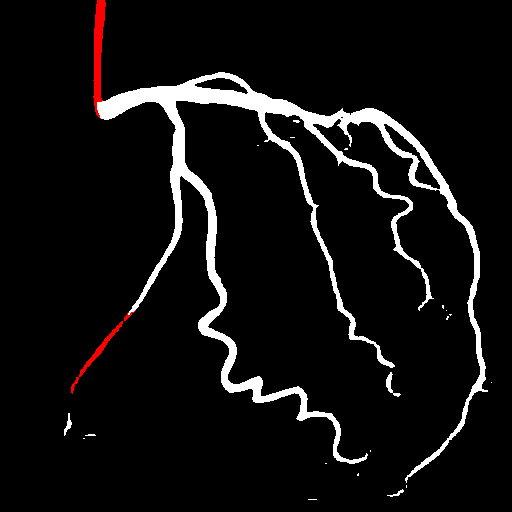}
        \caption{PAN}
        \label{fig:lca_pan}
    \end{subfigure}
    \begin{subfigure}[b]{0.24\linewidth}
        \centering
        \includegraphics[width=\linewidth]{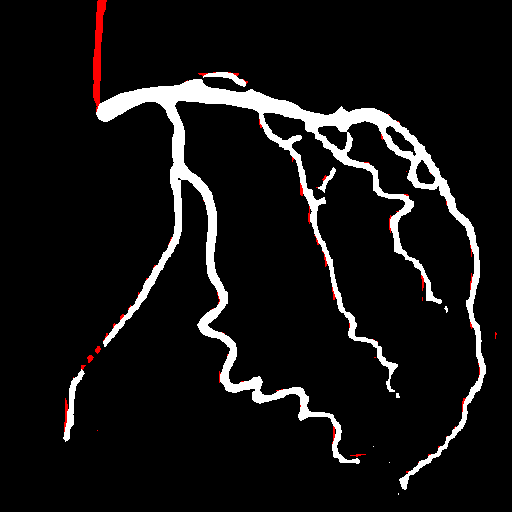}
        \caption{PSPNet}
        \label{fig:lca_pspnet}
    \end{subfigure}
    \caption{Segmentation of a left coronary artery. Each decoder was coupled with the encoder it performs best with. The masks are sorted in descending order of the models' average performances.}
    \label{fig:lca}
\end{figure}

\begin{figure}[h!]
    \centering
    \begin{subfigure}[b]{0.24\linewidth}
        \centering
        \includegraphics[width=\linewidth]{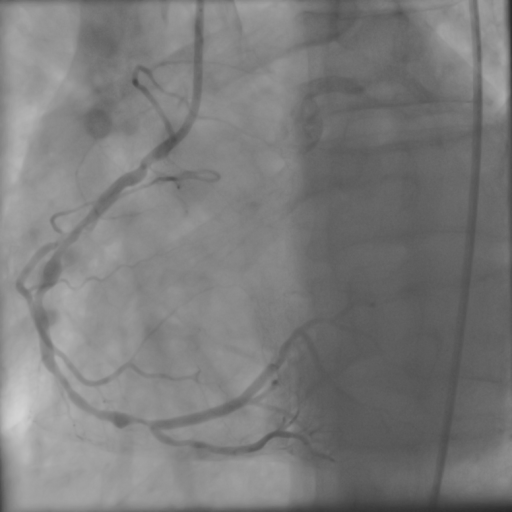}
        \caption{Angiogram}
        \label{fig:rca_img}
    \end{subfigure}
    \begin{subfigure}[b]{0.24\linewidth}
        \centering
        \includegraphics[width=\linewidth]{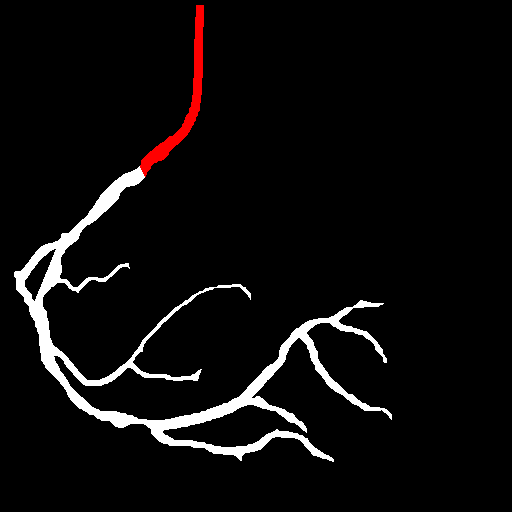}
        \caption{Ground-truth}
        \label{fig:rca_mask}
    \end{subfigure}
    \begin{subfigure}[b]{0.24\linewidth}
        \centering
        \includegraphics[width=\linewidth]{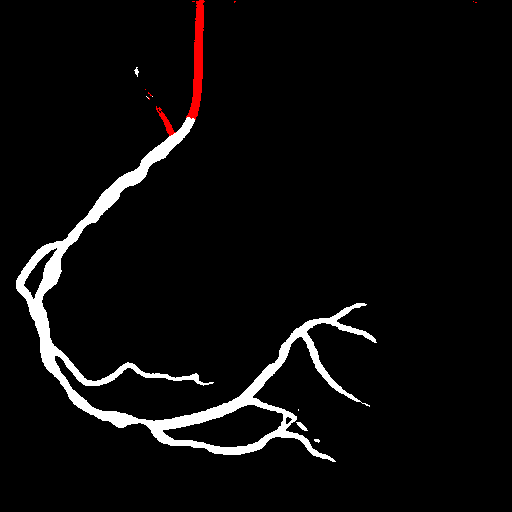}
        \caption{EfficientUNet++}
        \label{fig:rca_efficientunet++}
    \end{subfigure}
    \begin{subfigure}[b]{0.24\linewidth}
        \centering
        \includegraphics[width=\linewidth]{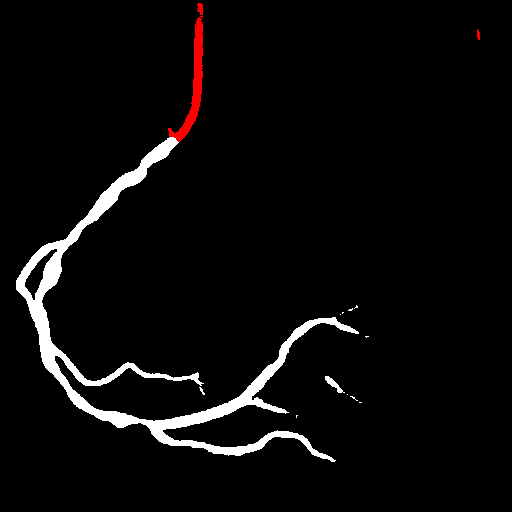}
        \caption{ResUNet++}
        \label{fig:rca_resunet++}
    \end{subfigure}
    \begin{subfigure}[b]{0.24\linewidth}
        \centering
        \includegraphics[width=\linewidth]{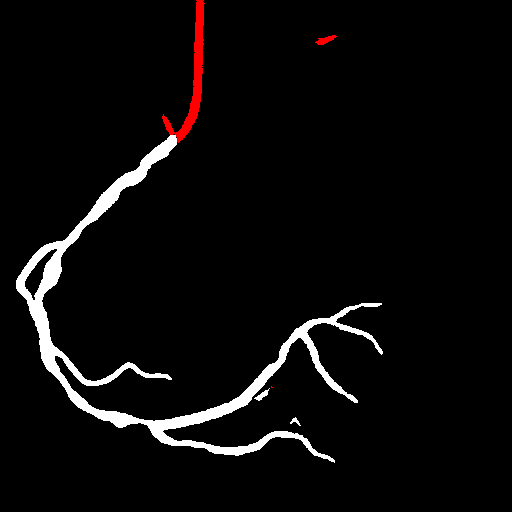}
        \caption{UNet++}
        \label{fig:rca_unet++}
    \end{subfigure}
    \begin{subfigure}[b]{0.24\linewidth}
        \centering
        \includegraphics[width=\linewidth]{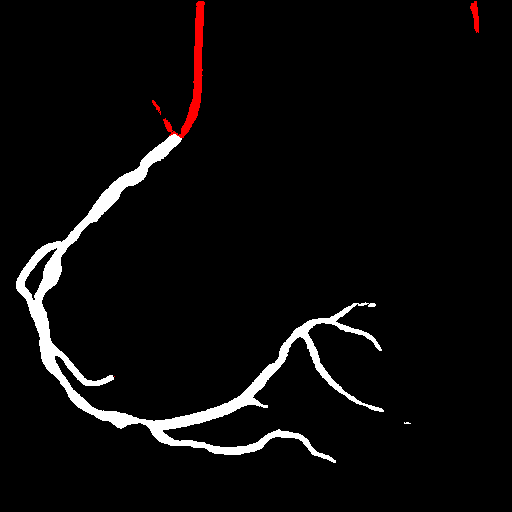}
        \caption{U-Net}
        \label{fig:rca_unet}
    \end{subfigure}
    \begin{subfigure}[b]{0.24\linewidth}
        \centering
        \includegraphics[width=\linewidth]{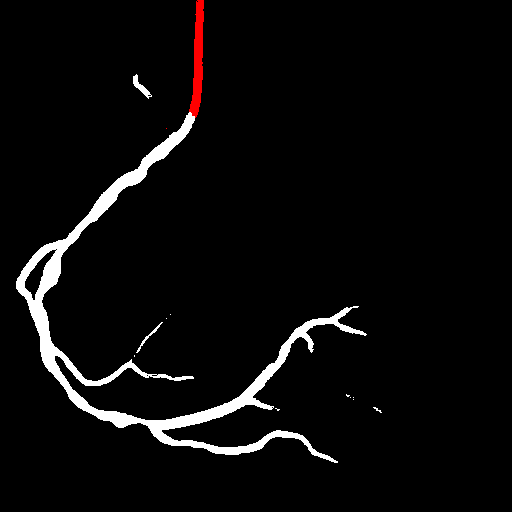}
        \caption{ResUNet}
        \label{fig:rca_resunet}
    \end{subfigure}
    \begin{subfigure}[b]{0.24\linewidth}
        \centering
        \includegraphics[width=\linewidth]{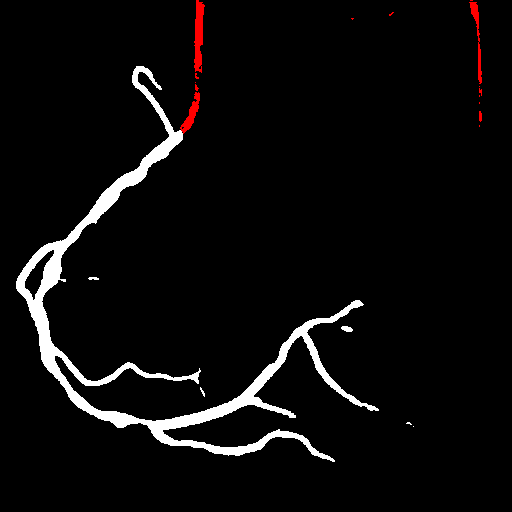}
        \caption{MANet}
        \label{fig:rca_manet}
    \end{subfigure}
    \begin{subfigure}[b]{0.24\linewidth}
        \centering
        \includegraphics[width=\linewidth]{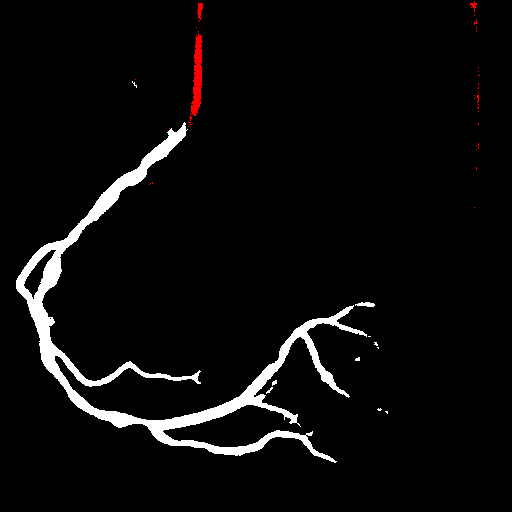}
        \caption{LinkNet}
        \label{fig:rca_linknet}
    \end{subfigure}
    \begin{subfigure}[b]{0.24\linewidth}
        \centering
        \includegraphics[width=\linewidth]{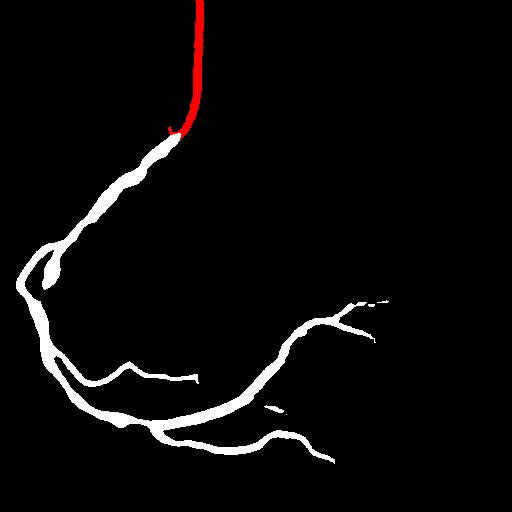}
        \caption{FPN}
        \label{fig:rca_fpn}
    \end{subfigure}
    \begin{subfigure}[b]{0.24\linewidth}
        \centering
        \includegraphics[width=\linewidth]{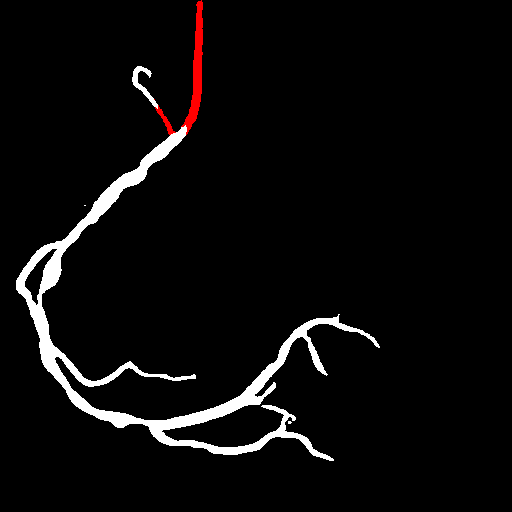}
        \caption{DeepLabV3+}
        \label{fig:rca_deeplabv3+}
    \end{subfigure}
    \begin{subfigure}[b]{0.24\linewidth}
        \centering
        \includegraphics[width=\linewidth]{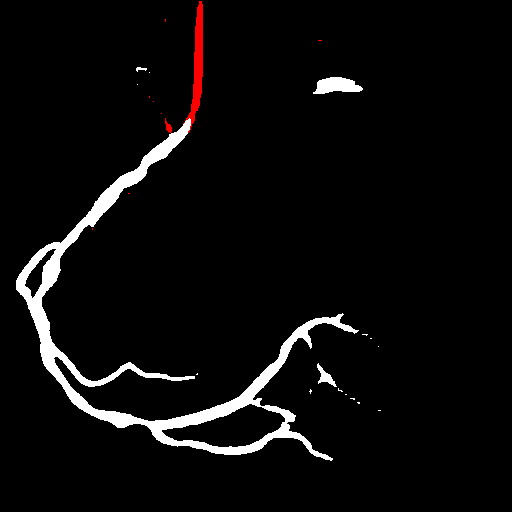}
        \caption{PAN}
        \label{fig:rca_pan}
    \end{subfigure}
    \begin{subfigure}[b]{0.24\linewidth}
        \centering
        \includegraphics[width=\linewidth]{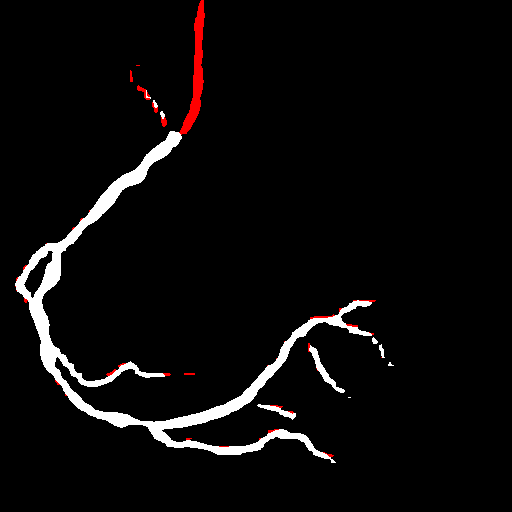}
        \caption{PSPNet}
        \label{fig:rca_pspnet}
    \end{subfigure}
    \caption{Segmentation of a right coronary artery. Each decoder was coupled with the encoder it performs best with. The masks are sorted in descending order of the models' average performances.}
    \label{fig:rca}
\end{figure}

Figures \ref{fig:lca} and \ref{fig:rca} show the segmentation masks obtained by each decoder when coupled with the encoder it performs best with, for a left and a right coronary artery, respectively. Visual inspection of these images supports the results of quantitative comparison. The U-Net-based models, except for the PAN, generally perform well. The FPN and DeepLabV3+ produce coarser yet very reasonable results, and the PSPNet performs very poorly, having trouble distinguishing catheters from arteries. 

The four best-performing architectures, i.e., the EfficientUNet++, ResUNet++, UNet++ and U-Net, output very good segmentation masks with few differences between them. Given the qualitative similarity between these models, the main advantage of the EfficientUNet++ is its higher computational efficiency. Also, we hypothesize that due to the attention mechanisms, it has more room for improvement than other architectures if trained on a larger dataset.

\clearpage

\section{Encoder and Decoder Performance Comparison Graphs}

\begin{figure}[h!]
    \centering
    \begin{subfigure}[b]{0.48\linewidth}
        \centering
        \includegraphics[width=\linewidth]{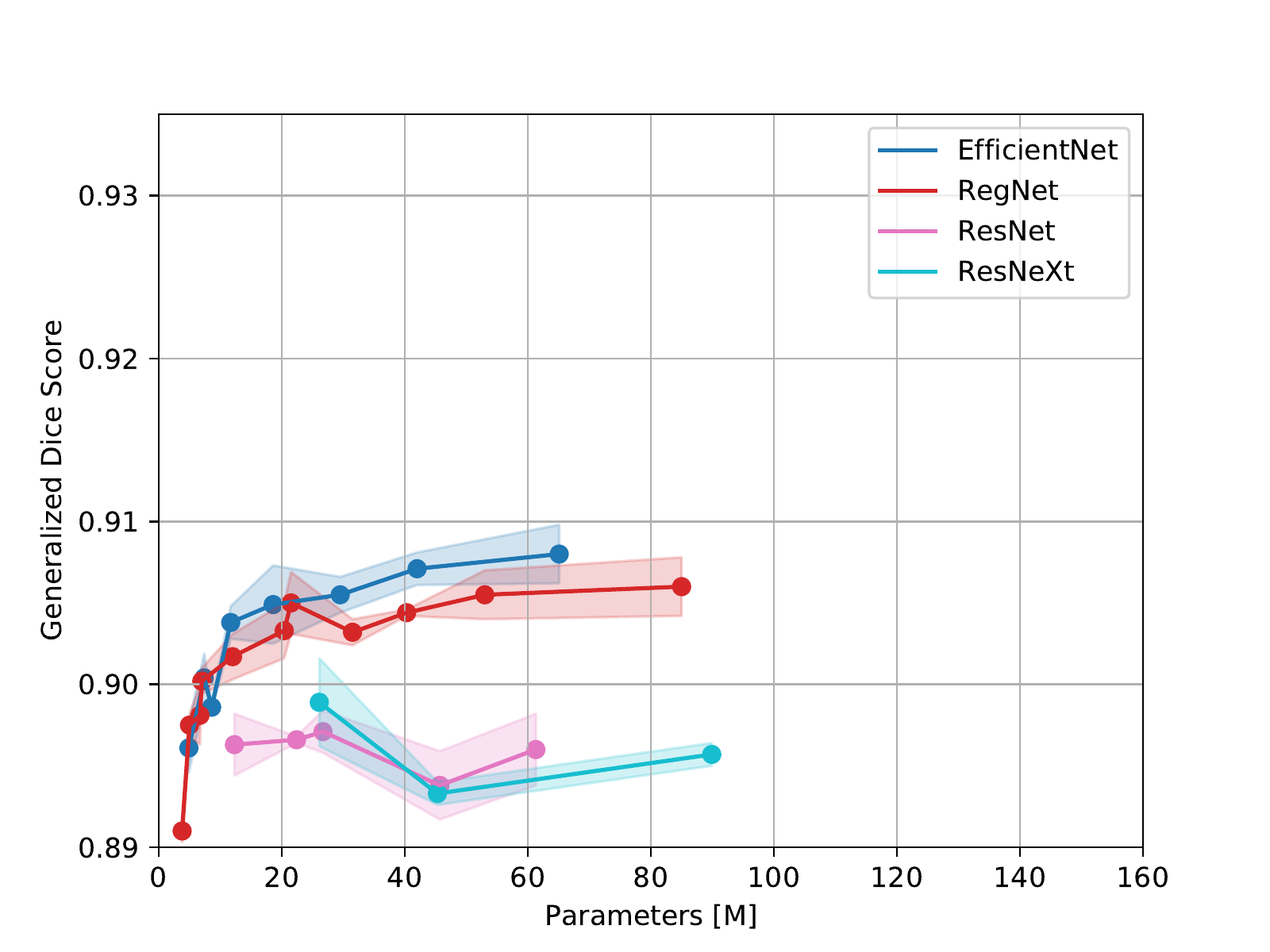}
        \caption{DeepLabV3+ decoder}
        \label{fig:deeplabv3+params}
    \end{subfigure}
    \begin{subfigure}[b]{0.48\linewidth}
        \centering
        \includegraphics[width=\linewidth]{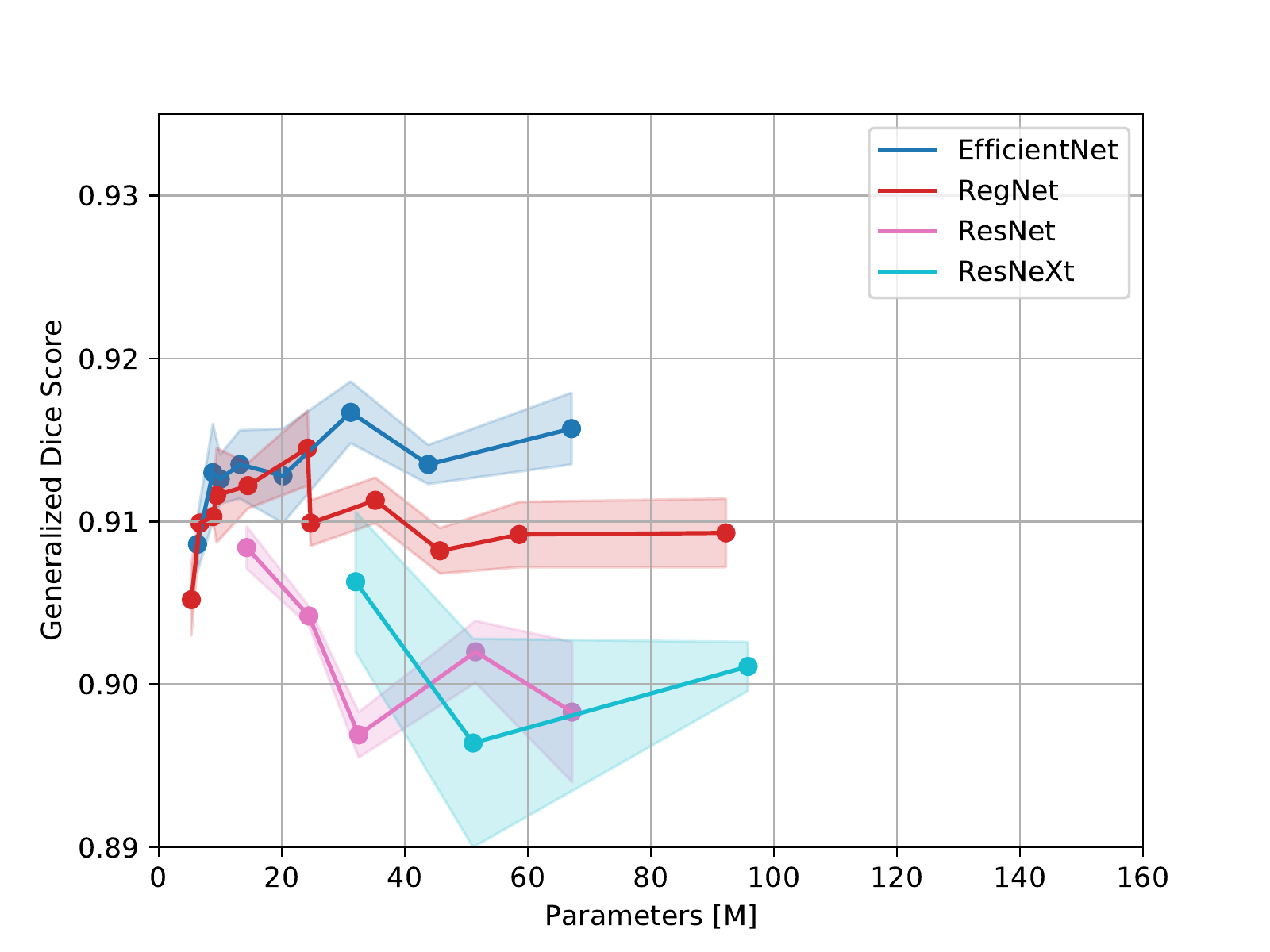}
        \caption{U-Net decoder}
        \label{fig:unetparams}
    \end{subfigure}
    \begin{subfigure}[b]{0.48\linewidth}
        \centering
        \includegraphics[width=\linewidth]{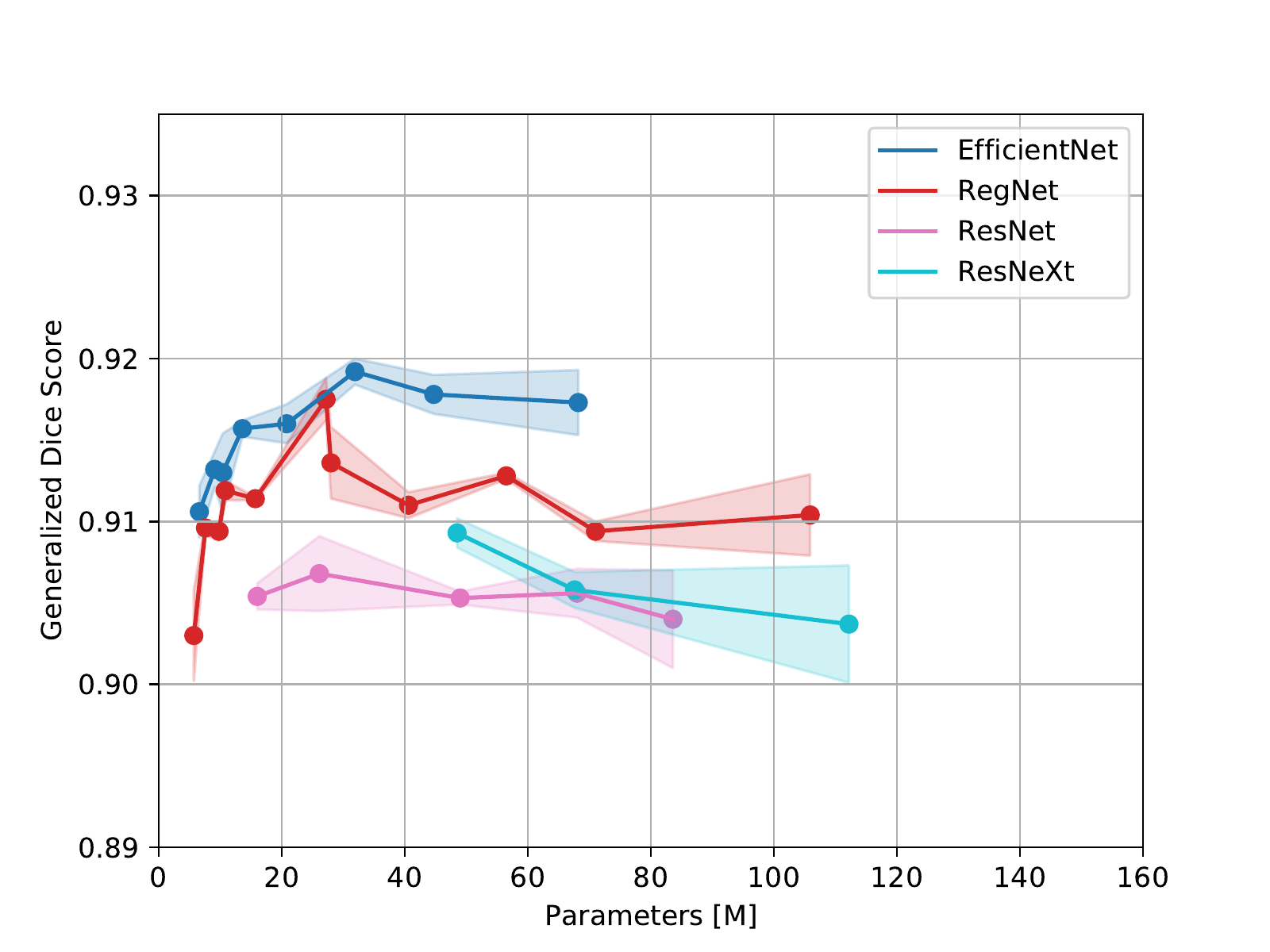}
        \caption{UNet++ decoder}
        \label{fig:unet++params}
    \end{subfigure}
    \begin{subfigure}[b]{0.48\linewidth}
        \centering
        \includegraphics[width=\linewidth]{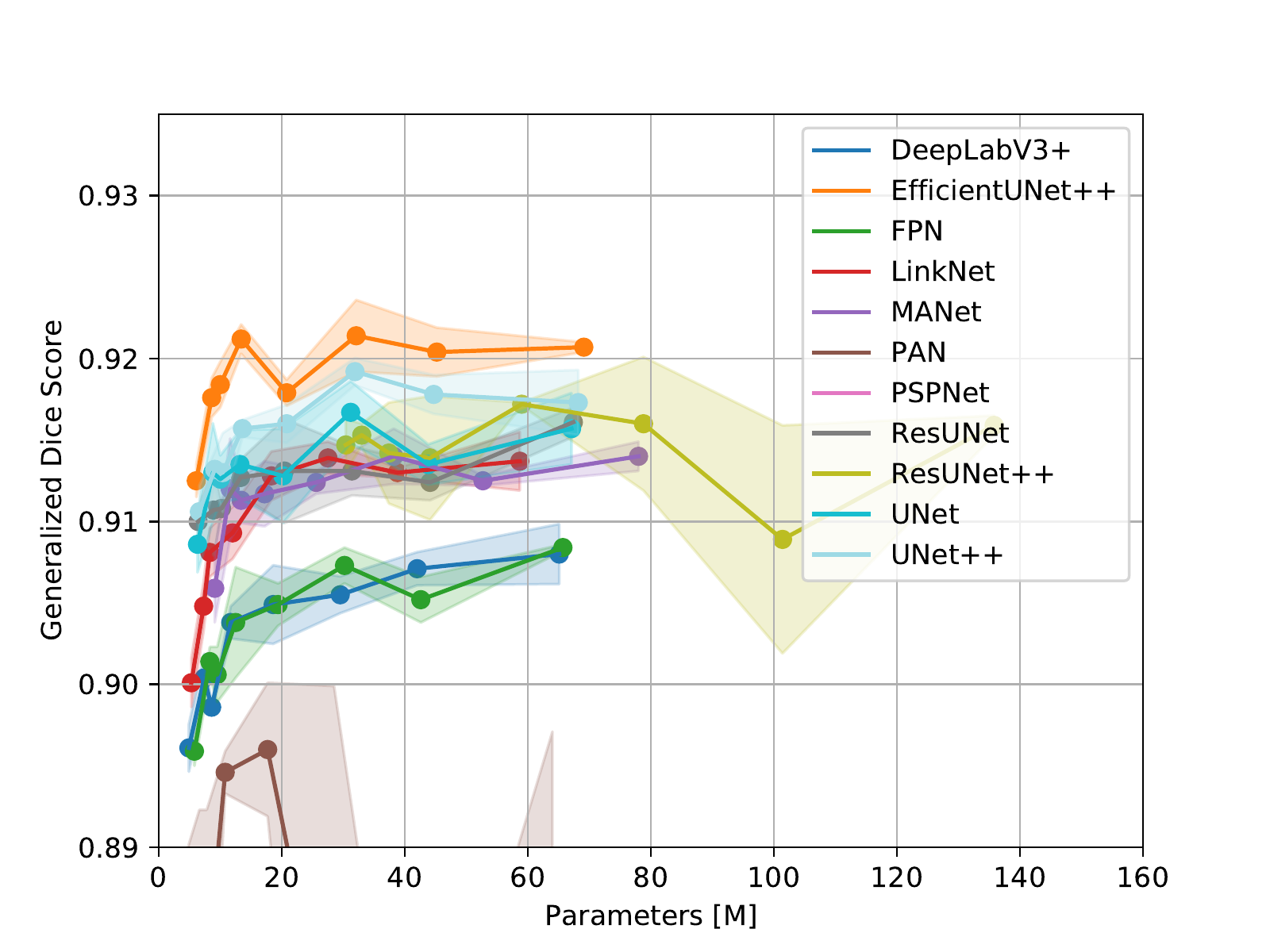}
        \caption{EfficientNet B0 to B7 encoders}
        \label{fig:decoder_params}
    \end{subfigure}
    \caption{Segmentation performance, measured by generalized dice score (GDS), as a function of the number of parameters. Figures (a), (b) and (c) show the performance of different encoders combined with the (a) DeepLabV3+, (b) U-Net and (c) UNet++ decoders. Figure (d) shows the performance of different decoders combined with the EfficientNet B0 to B7 encoders. Each polygonal line corresponds to an encoder family. The dots represent the following models, in ascending order of parameters: EfficientNet - B0, B1, B2, B3, B4, B5, B6, B7; RegNet - Y2, Y4, Y6, Y8, Y16, Y32, Y40, Y64, Y80, Y120, Y160; ResNet - 18, 34, 50, 101, 152; ResNeXt - 50\_32x4d, 101\_32x4d, 101\_32x8d. Models with GDS below 0.89 are also omitted.}
    \label{fig:comparison_params}
\end{figure}